\documentclass[epj]{svjour}
\usepackage{graphicx}
\usepackage{mathtext}
\usepackage{amsmath}

\begin{document}

\title{Boiling of the Interface between Two Immiscible Liquids
       below the Bulk Boiling Temperatures of Both Components}
\titlerunning{Boiling of the interface between immiscible liquids}

\author{Anastasiya~V.~Pimenova\inst{1}
        \and Denis~S.~Goldobin\inst{1,2,3}
        }
\authorrunning{A.~V.~Pimenova, D.~S.~Goldobin}

\institute{Institute of Continuous Media Mechanics, UB RAS,
 Perm 614013, Russia
\and Department of Mathematics, University of Leicester,
 Leicester LE1 7RH, UK
\and Perm State National Research University,
 Perm 614990, Russia
}
\date{\today}

\abstract{
We consider the problem of boiling of the direct contact of two
immiscible liquids. An intense vapour formation at such a direct
contact is possible below the bulk boiling points of both
components, meaning an effective decrease of the boiling
temperature of the system. Although the phenomenon is known in
science and widely employed in technology, the direct contact
boiling process was thoroughly studied (both experimentally and
theoretically) only for the case where one of liquids is becoming
heated above its bulk boiling point. On the contrary, we address
the case where both liquids remain below their bulk boiling
points. In this paper we construct the theoretical description of
the boiling process and discuss the actualisation of the case we
consider for real systems.
}

\maketitle

\section{Introduction}
\label{sec:intro}
The process of boiling of a system of two immiscible liquids has a
remarkable feature: it can occur at temperatures below the bulk
boiling temperatures of both components (e.g.,
see~\cite{Krell-1982,Geankoplis-2003}). This phenomenon is
explained by the fact that boiling occurs at the interface between
two liquids, but not in their bulk. Molecules from both liquids
evaporate into the vapour layer between two liquids and each
liquid tends to be in local thermodynamic equilibrium with its
vapour, therefore equilibrium pressure within this layer is equal
to the sum of the saturated vapour pressures of both liquids.
Hence, the condition for the growth of the vapour phase is the
exceeding of atmospheric pressure by the sum of the saturated
vapour pressures, while for the bulk boiling the vapour pressure
alone should exceed atmospheric pressure.

The phenomenon under consideration is widely used in
industry~\cite{Krell-1982,Geankoplis-2003}. For instance, it is
beneficial for distillation of substances the boiling temperature
of which is higher than the decomposition temperature at
atmospheric pressure (as it is for insoluble tetraethyllead). This
phenomenon is employed for combustion of poorly volatile liquid
fuels in furnaces. Simultaneously, it is the reason why water is
forbidden for usage when one needs to stop fire of inflammable
organic liquids. The phenomenon is also of interest in relation to
the process of combustion of a light inflammable liquid covering
the surface of a heavy nonflammable liquid.

Although this phenomenon is well known in the literature, many
experimental~\cite{Simpson-etal-1974,Celata-1995,Roesle-Kulacki-2012-1,Roesle-Kulacki-2012-2}
and theoretical works~\cite{Sideman-Isenberg-1967,Kendoush-2004}
deal with the case where one of components is heated above its
bulk boiling temperature. The case of interfacial boiling below
the bulk boiling temperatures of both components did not receive a
thorough study in the literature. Meanwhile, this case is most
intriguing as the one where boiling becomes possible being
impossible otherwise. Moreover, there are situations in real
systems, where exactly this case becomes relevant (see
Sec.~\ref{sec:example}).

The interfacial boiling starts at temperature $T_\ast$ determined
by the condition that the consolidated pressure of the saturated
vapours of both liquids is equal to atmospheric pressure. In terms
of the particle number densities $n_j^{(0)}$ of saturated vapours:
\begin{equation}
n_1^{(0)}(T_\ast)+n_1^{(0)}(T_\ast)=\frac{p_0}{k_\mathrm{B}T_\ast}\,,
\nonumber
\end{equation}
where $p_0$ is atmospheric pressure, $k_\mathrm{B}$ is the
Boltzmann constant. We will consider the case where both
components are below their bulk boiling points, i.e., the
temperature field in the system does not exceed $T_\ast$
significantly. In this case, vapour is generated only at the
direct contact of two liquid; a growing vapour layer forms in
between the liquids and experiences ``resetting'' to zero
thickness from time to time because of vapour breakaway.

\begin{figure*}[t]
\begin{tabular}{ccc}
\hspace{-7pt}
\includegraphics[width=0.3205\textwidth]%
{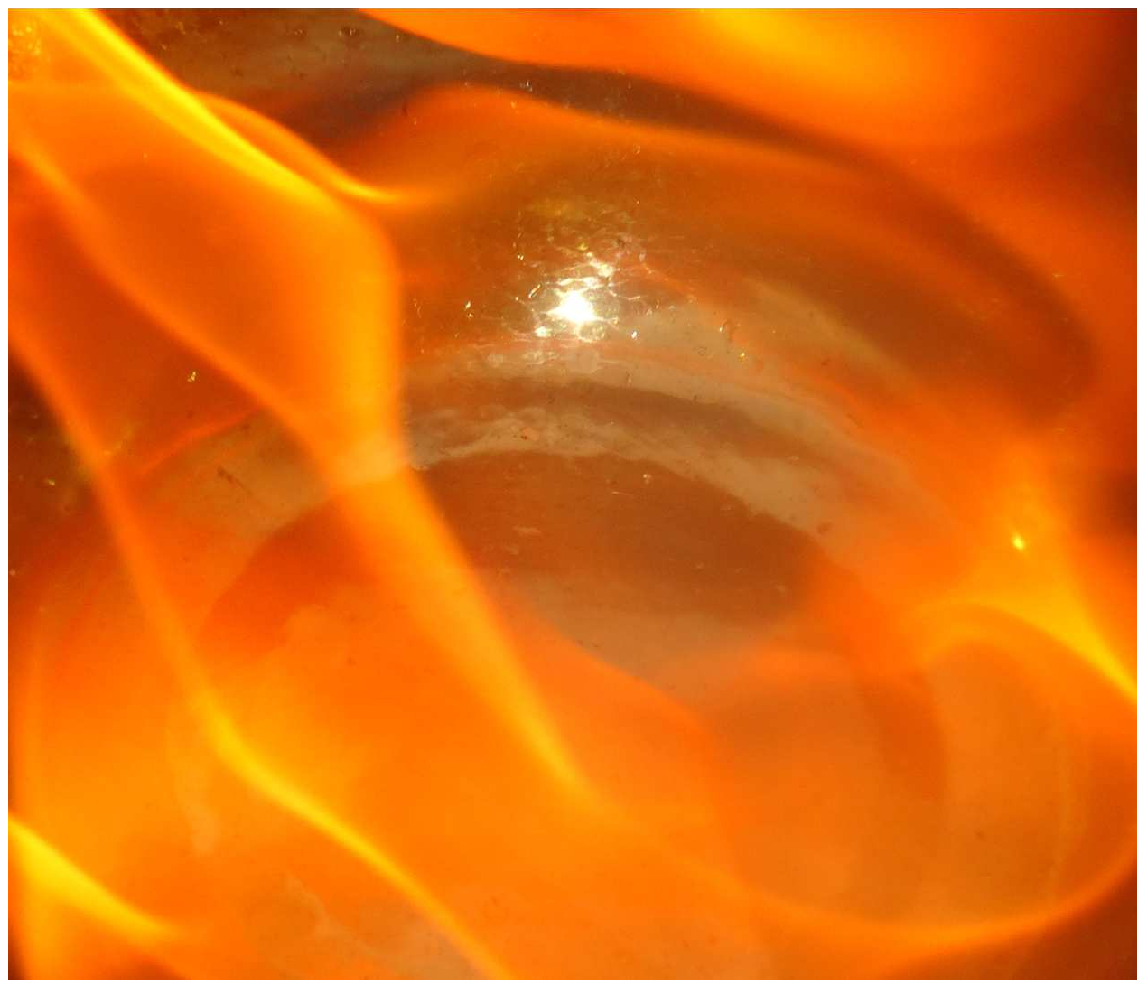}
\hspace{-7pt}
 &
\hspace{-7pt}
\includegraphics[width=0.3205\textwidth]%
{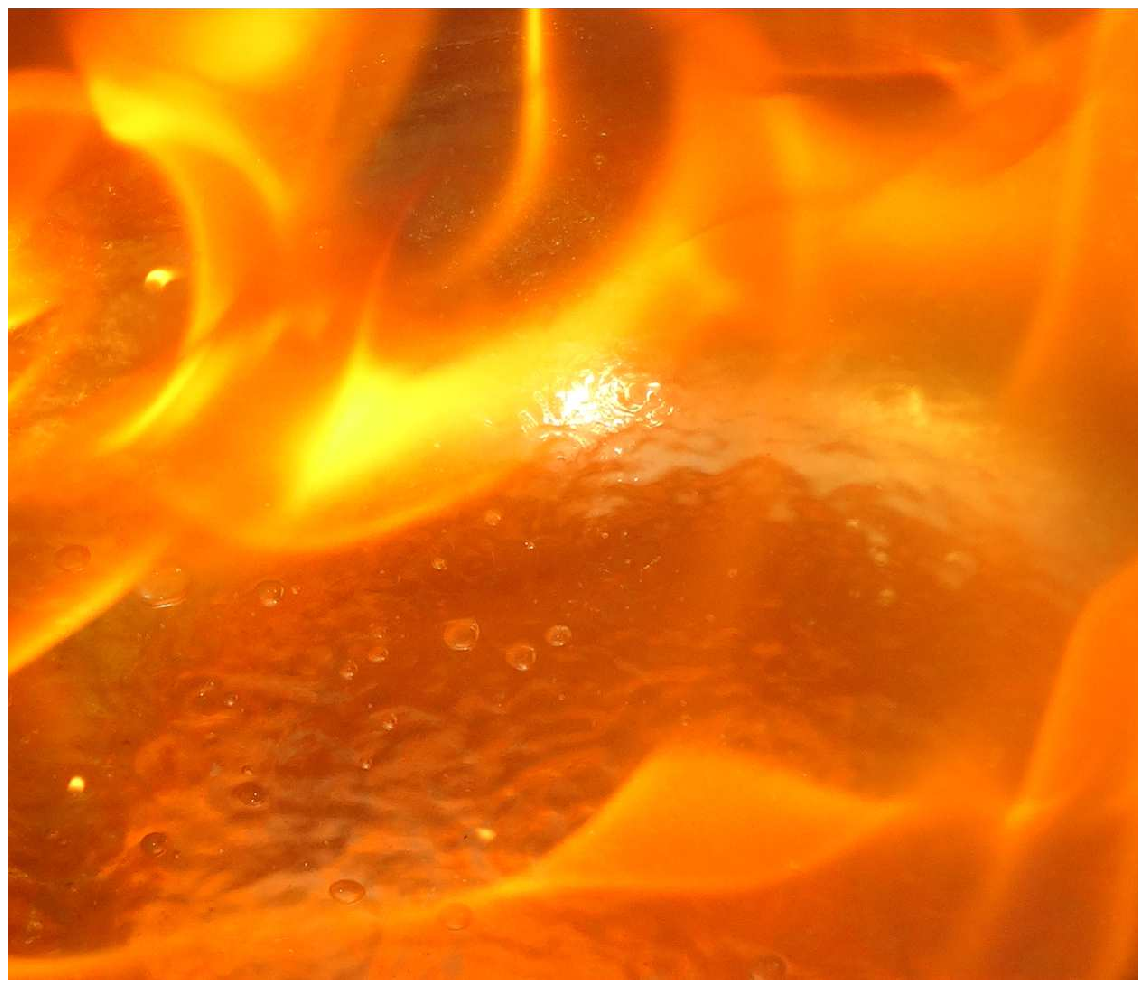}
\hspace{-7pt}
 &
\hspace{-7pt}
\includegraphics[width=0.3205\textwidth]%
{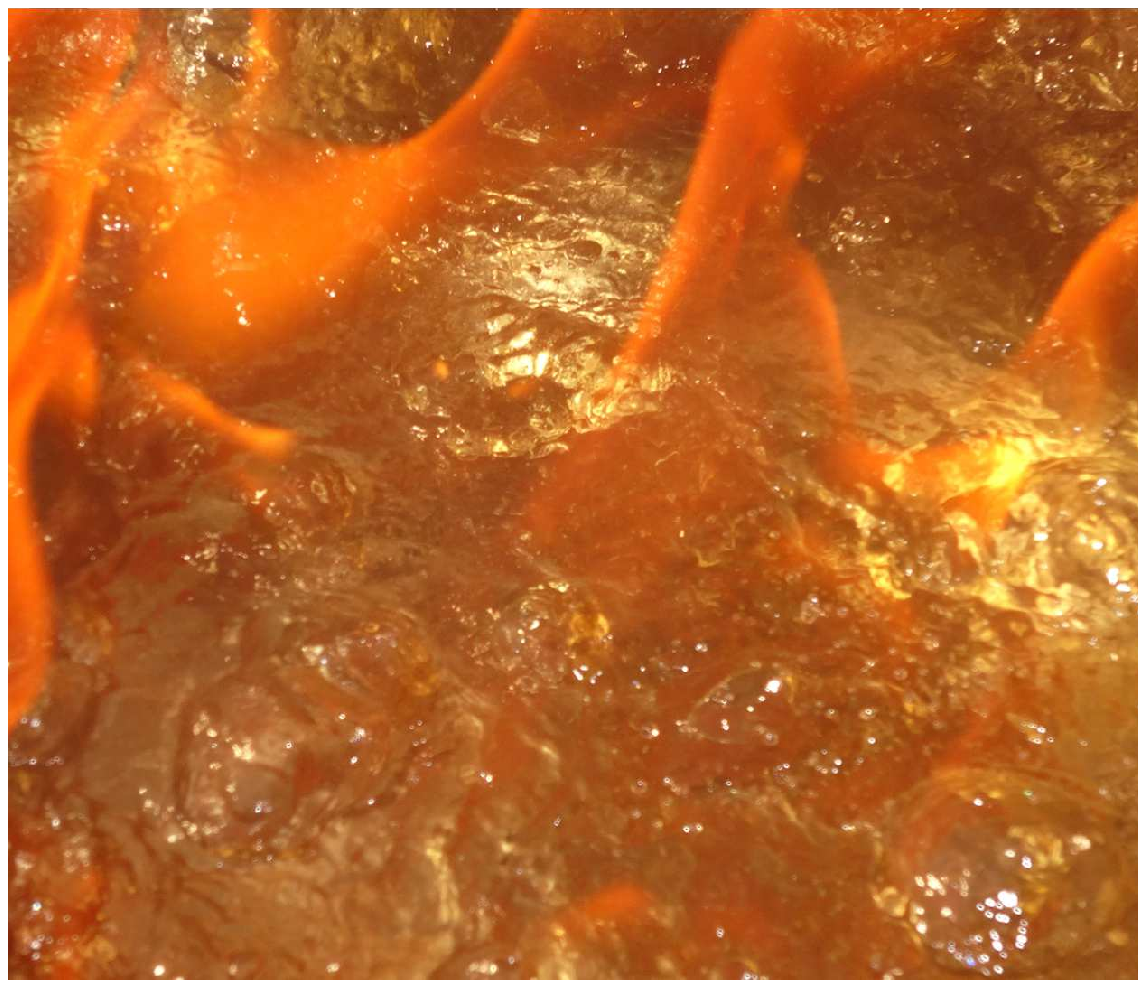}
\hspace{-7pt}
\\
{\sf (a)} & {\sf (b)} & {\sf (c)}
\end{tabular}

\caption{
Demonstration experiment: Combustion of layer of ``white spirit''
(light fractions of kerosine) over water. (a):~Nearly immediately
after ignition, the flammable liquid and water are stably
stratified (convective flows are suppressed); there is only
intense surface evaporation of burning liquid, no bulk boiling.
This regime is the regime of combustion as it would occur without
underlying liquid. (b):~$2\,\mathrm{min}$ after ignition, one can
see rare vapour bubbles rising from the white spirit--water
interface. The bulk boiling of water does not occur, meaning the
interface temperature is below the bulk boiling points of both
liquids. Intensity of interface boiling increases over time as
white spirit layer becomes thinner and the heat influx to the
interface increases. (c):~$4\,\mathrm{min}$ after ignition,
boiling of the interface is intense, although a stratified state
persists in some form. {\bf Demonstration experiment set-up}: The
two-layer system was placed in a steel cavity of internal diameter
$16\,\mathrm{cm}$; the initial thicknesses of white spirit and
water layers: $2\,\mathrm{cm}$ and $3\,\mathrm{cm}$, respectively;
air temperature and initial temperature of white spirit:
$20^\circ\mathrm{C}$, initial water temperature:
$40^\circ\mathrm{C}$. Timing for the combustion process
significantly varies depending on air temperature and wind
strength.
 }
\label{fig1}
\end{figure*}

For methodological reasons, the particular case of two liquids
with nearly identical values of physical chemical parameters was
addressed as a first step of theoretical study of the
phenomenon~\cite{Pimenova-Goldobin-JETP-2014}. In this paper we
extend the consideration to the case of different quantitative
characteristics of liquids and also allow for the asymmetry
between states of two liquids (which is important even when their
properties are similar). Additionally, the demonstration
experiments with combustion of a light inflammable liquid layer
over a heavy nonflammable one are described and an auxiliary
problem of the hydrodynamic instability of a thin vapour layer
between heavy and light liquids to bubble formation is
investigated in the Appendix.

The paper is organised as follows. In Sec.~\ref{sec:example}, the
process of combustion of a light inflammable liquid over a heavy
nonflammable one is discussed. In Sec.~\ref{sec:model}, we
formulate the specific physical problem we deal with and derive
the mathematical model of the system from scratch. In
Sec.~\ref{sec:layer_growth} we derive the solution to the
mathematical model, which describes the growth of the vapour
layer. In Sec.~\ref{sec:macroscopic}, relationships between
macroscopic quantifiers of the system state and the derived
growing-vapour-layer solution are established; the problem of the
vapour bubble formation and the associated vapour layer breakaway
is addressed for the cases of a well-stirred system and a
stratified one. In Sec.~\ref{sec:simplific}, we overview the
simplification assumptions of our work and perform quantitative
assessments related to their accuracy. In
Sec.~\ref{sec:conclusion}, we draw conclusions. Solutions to
several auxiliary problems are provided in Appendices.

\begin{figure}[t]
\begin{tabular}{c}
\hspace{-1.25mm}{\sf (a)}
\includegraphics[width=0.445\textwidth]%
{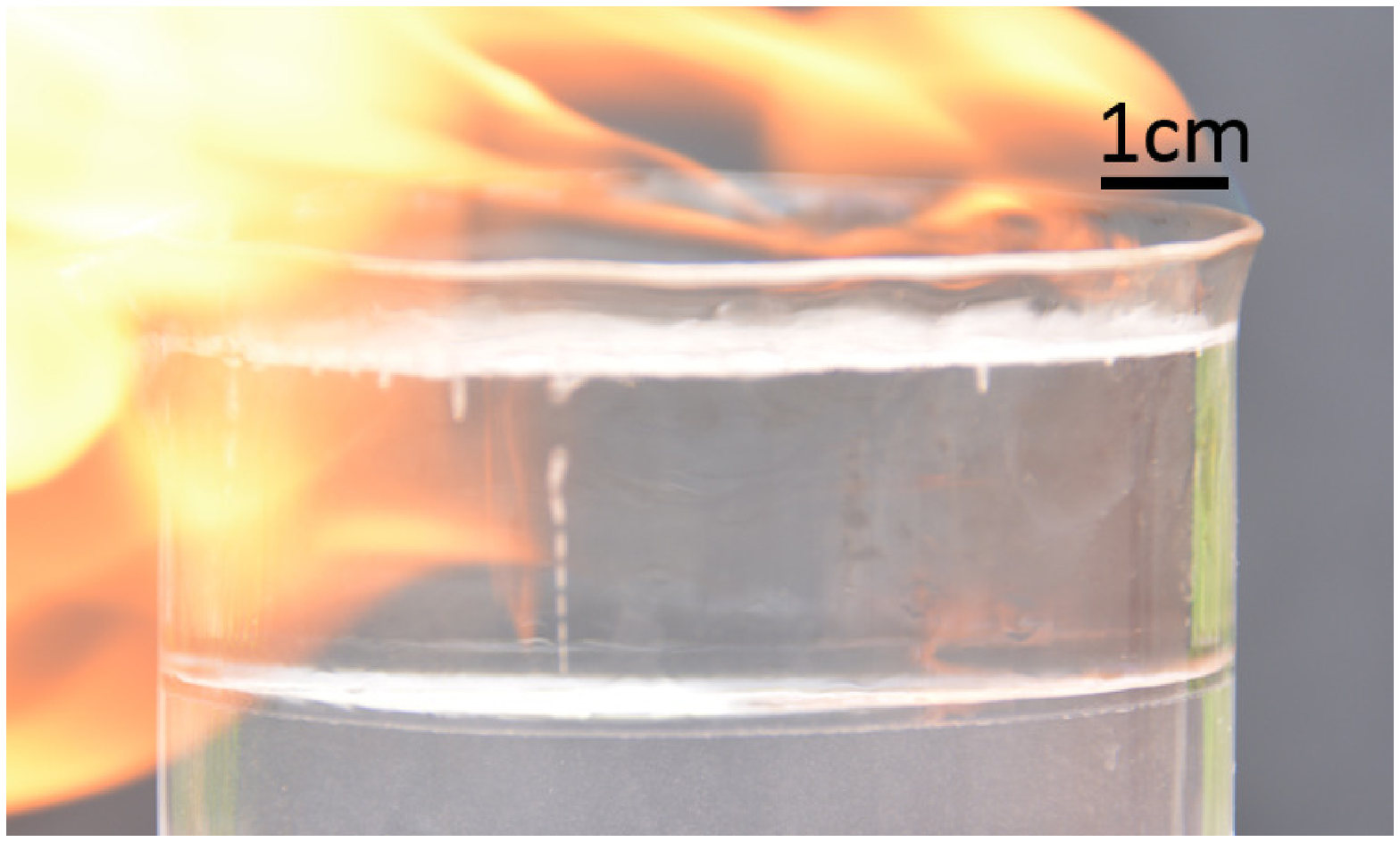}
\\[15pt]
\hspace{-1.25mm}{\sf (b)}
\includegraphics[width=0.445\textwidth]%
{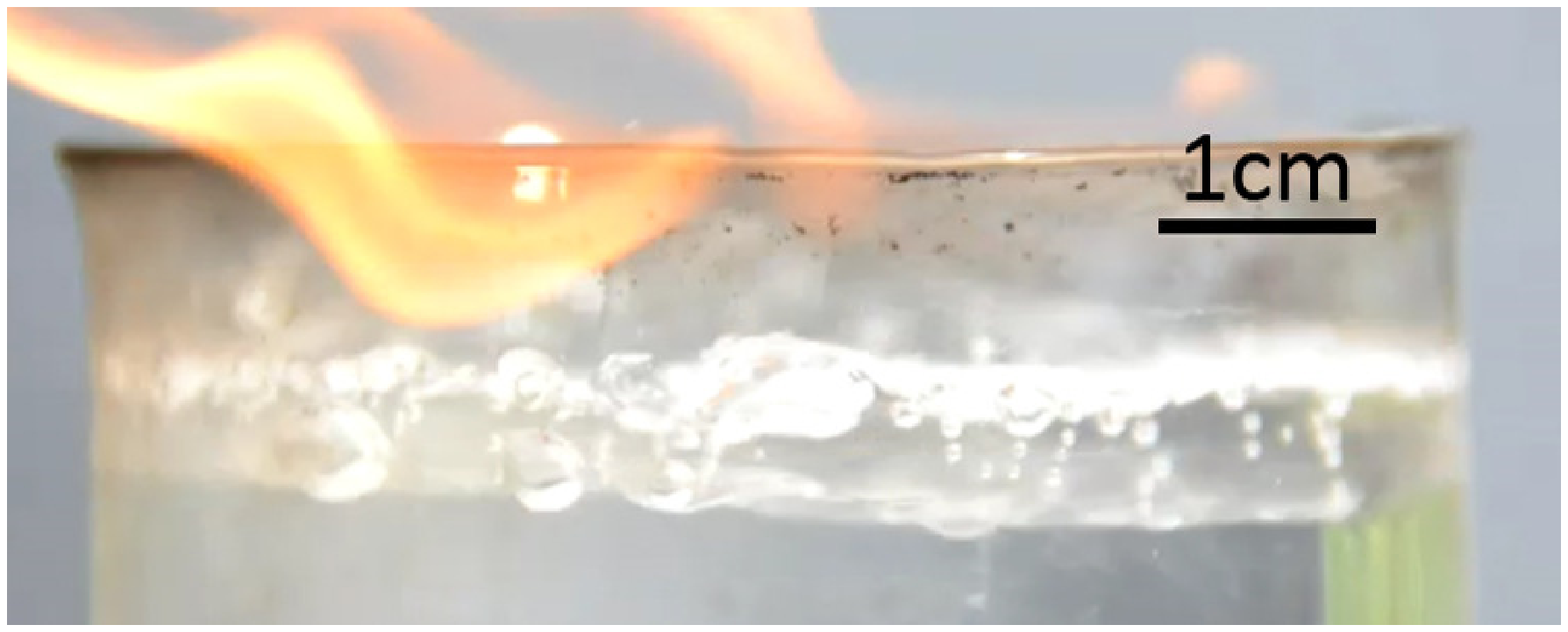}
\\[5pt]
\end{tabular}

\caption{
Demonstration experiment: Combustion of layer of n-heptane over
water. (a)~Early stage of boiling of the h-heptane--water
interface: one centre of vapour formation and a bubble lane from
it can be observed. (b):~Intense boiling at the interface: a
plenty of bubble lanes running from the interface can be observed,
nearly all of them are away from the glass wall. {\bf
Demonstration experiment set-up}: The two-layer system was placed
in a quartz glass of diameter $7\,\mathrm{cm}$; air temperature
and initial temperature of n-heptane: $20^\circ\mathrm{C}$,
initial water temperature: $90^\circ\mathrm{C}$.
 }
\label{fig2}
\end{figure}

\section{Example: Combustion of light inflammable liquid over heavy nonflammable one}
\label{sec:example}
Prior to constructing the phenomenon theory, we would like to
substantiate our interest to the specific case we consider not
only by the reason of academic curiosity and non-triviality of the
phenomenon of the decrease of the boiling point but also by
practical reasons. We intend to discuss the primary relevance of
specifically the case under consideration for combustion of a
light inflammable liquid over a heavy nonflammable one.

Necessity of this argumentation is dictated by the fact that
previously only the case of superheating conditions for one of
components was addressed in experimental studies. This choice for
experiment setups was related to industrial applications. However,
we are to explain that this is not the only case which can be of
practical importance. Obviously, even for the case, where one of
components to be mainly superheated, the system unavoidably passes
through the transient regimes where interfacial boiling still or
already occurs but both components are not superheated. These can
be late stages of self-cooling of the system without heat supply
or early stages of mixing of two liquids, temperatures of which
are such that one of liquids will be superheated before the system
reaches thermal equilibrium. However, there are situations where
the system is maintaining itself in the regime of our interest,
the regime persists but not occur as a stage of a transient
process. For these situations the relevance of our work is more
pronounced.

Let us consider combustion of a light inflammable liquid over a
heavy nonflammable one. We need first to remind that combustion of
an inflammable liquid in an open container happens without the
bulk boiling: there is only an intense surface evaporation.
Indeed, the burning surface is heated to the bulk boiling
temperature and the bulk of liquid is at lower temperature. If the
heat influx from the flame becomes strong enough to heat the bulk
above the boiling point, intense vapour formation starts. Given
not enough oxygen (which is accessible only outside the liquid)
provided, the vapour combustion area (flame) will be pushed away
from the liquid by an intense vapour flux and thus liquid heating
efficiency will be decreased.[\footnote{Alternatively, given there
is enough oxygen for immediate combustion of the excessive vapour,
the combustion process will become explosive. As long as there is
no explosion, one can surely conclude that the combustion occurs
without bulk boiling.}] Noteworthy, in the course of such a
combustion the liquid is stably stratified due to temperature
gradient and all convective currents are
suppressed~\cite{Incropera-etal-2006}. The presence of another
liquid below the burning one can change the situation. The
interface between these two liquids can be heated to the
interfacial boiling temperature without exceeding the bulk boiling
temperature within the burning liquid. Boiling of this interface
may have different effects on the combustion process, depending on
the physical properties of liquids and system configuration.
However, it is certain, that the interfacial boiling will be
maintained at conditions we are to consider. Moreover, if the
inflammable liquid is more volatile than the heavier bottom liquid
(for instance, n-heptane over water), both components will be
never superheated.

We performed demonstration experiments with ``white
spirit''--water (Fig.~\ref{fig1}) and n-heptane--water
(Fig.~\ref{fig2}) systems. With these demonstration experiments
(see Figs.~\ref{fig1}, \ref{fig2} and video in supplementary
material or on YouTube~\cite{YouTube}) one can observe the
interfacial boiling in the course of combustion of organic fuel.
The effect of this interfacial boiling on the combustion process
will be considered elsewhere in detail; here we provide this
example in support of physical and practical relevance of the
problem set-up we will be using.

\begin{figure}[t]
\center{
\includegraphics[width=0.485\textwidth]%
{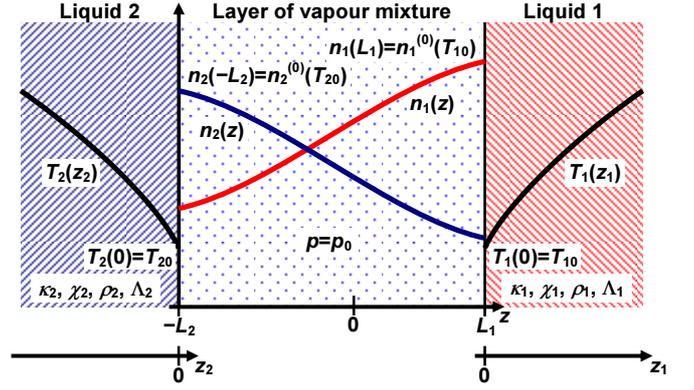}}
\caption{
Growing vapour layer between two half-spaces of immiscible liquids
and the reference frames.}
\label{fig3}
\end{figure}

\section{Evolution of the vapour layer}
\label{sec:model}
The mathematical description of the dynamics of the system
(Fig.~\ref{fig3}) is based on the following physical assumptions:
\begin{enumerate}
\item
Temperature in the zones of the liquid phases is  nonuniform in
order to provide the heat inflow to the surface of the
evaporation.
\item
The mass of the liquid phase decreases in the course of
evaporation and the surfaces move deeper into liquids.
\item
The substances evaporate from the surfaces of liquids into the
vapour layer. We consider liquids which are mutually insoluble;
therefore, the molecules of the first substance do not pass into
the liquid phase at the second liquid--vapour interface.
Immediately above the liquid surface, the number density $n_j$
($j=1,2$) of the particles of the corresponding substance is equal
to the particle number density of the saturated vapour, say
$n_j^{(0)}$, of this substance at local temperature $T$
\[
n_2(z=-L_2)=n_2^{(0)}(T),\quad
n_1(z=L_1)=n_1^{(0)}(T).
\]
When the liquid evaporate from free surface (e.g.\ into vacuum),
the number density of the vapour above the surface is lower than
the one of the saturated vapour due to finite escape rate of the
evaporant and rapid diffusion outflow of particles from the
surface~\cite{Anisimov-Rakhmatulina-1973}. In the system under
consideration, liquids evaporate not into the open half-space, but
into the vapour layer. The differences of the number densities
across the layer tend to zero for $T\to T_\ast$, and the diffusion
outflow becomes asymptotically small, making the transition rates
between the vapour and liquid phases sufficient to maintain the
local thermodynamic equilibrium between the phases. Thus, the
assumption of the local thermodynamic equilibrium at the
vapour-liquid interface can be considered to be valid by
continuity as long as the overheating of the system is small
enough.
\item
Pressure within the vapour layer is assumed to be constant and
equal to atmospheric pressure, $p=p_0$, due to the
negligibility of the mechanical inertia compared to the heat and
diffusion ``inertia'' of the system.
\item
The total number density of particles in the vapour layer obeys
the ideal gas law
\[
n_1+n_2=n_0=\frac{p_0}{k_B T}\,,
\]
where $T$ is local temperature. The variation of $n_0$ associated
with the temperature deviation from $T_\ast$ is negligible
compared to the variation of $n_j$; thus,
\begin{equation}
n_1+n_2=n_{0\ast}=\frac{p_0}{k_B T_\ast}\,.
\label{eq-01}
\end{equation}
\end{enumerate}

For the interfacial boiling to occur the system must be overheated
above the minimal temperature of boiling $T_\ast$:
$$
T=T_\ast+\Theta\,,\qquad\Theta>0\,.
$$
We focus our consideration on the case where both liquids are far
from boiling conditions in their bulk, as the most intriguing one
from the research view point. For the vapour layer, where heat is
consumed for evaporation, the temperature excess above $T_\ast$ is
even smaller than for the bulk. Hence, one can linearise the
dependence of the saturated vapour number density $n_j^{(0)}(T)$
about $T=T_\ast$;
\begin{equation}
n_j^{(0)}(T_\ast+\Theta)=n_{j\ast}^{(0)}+\gamma_j\Theta+\dots\,,
\label{eq-02}
\end{equation}
where $n_{j\ast}^{(0)}\equiv n_j^{(0)}(T_\ast)$ and
$\gamma_j\equiv(\partial n_j^{(0)}/\partial T)_{T=T_\ast}$.

We deal with temperature variations which are small compared
to the absolute temperature, $\delta T/T\ll1$; therefore, for physical
parameters which depend on temperature polynomially one can
neglect variations due to their smallness (e.g., $n_0\propto 1/T$
and $|\delta n_0|/n_0=|\delta T|/T\ll1$). On the contrary, for
parameters which are exponential in $T$ one has to take this
dependence into account. E.g., for the number density of saturated
vapour
 $n^{(0)}(T)/n^{(0)}(T_\ast)\propto e^{(T-T_\ast)/T_0}$,
where $T_0$ is of the order of magnitude of $10\,\mathrm{K}$, and
 $|\delta n^{(0)}|/n^{(0)}=|\delta T|/T_0$,
which is by factor 30--40 larger than $|\delta T|/T$. We will
construct our theory for small but non-negligible terms
$\sim|\delta T|/T_0$ and neglect contributions $\sim|\delta T|/T$.

It is convenient to use different reference frames for three
areas: for the liquids 1 and 2 the coordinates $z_1$ and $z_2$
measure shifts from the respective liquid surfaces and for the
vapour layer the coordinate $z$ is in the range from $-L_2$ to
$L_1$ (see Fig.~\ref{fig3}).

Let us first consider the molecule number balance in the system.
The number density redistribution of species within the vapour
layer is due to molecular diffusion governed by the Fick's law
\begin{equation}
\vec{J}_j =-D_{12} \nabla n_j,
\label{eq-03}
\end{equation}
where $D_{12}$ is the coefficient of mutual diffusion. The
dependence of this coefficient on temperature is a power law and,
therefore, can be neglected as discussed above. Thus, the
evolution of the number densities is governed by
\begin{equation}
\frac{\partial n_j}{\partial t}=\nabla\cdot(D_{12}\nabla n_j)
 =D_{12}\frac{\partial^2n_j}{\partial z^2}\,.
\label{eq-04}
\end{equation}

For the boundary conditions on the number densities, we employ
linearised dependencies $n_j^{(0)}(T)$, Eq.~(\ref{eq-02});
\begin{align}
n_1|_{z=L_1}&=n_{1*}^{(0)}+\gamma_1\Theta\,,
\label{eq-05}\\
n_2|_{z=-L_2}&=n_{2*}^{(0)}+\gamma_2\Theta\,.
\label{eq-06}
\end{align}
Due to Eq.~(\ref{eq-01}), Eqs.~(\ref{eq-05})--(\ref{eq-06}) yield
as well
\begin{align}
n_2|_{z=L_1}&=n_{2*}^{(0)}-\gamma_1\Theta\,,
\label{eq-07}\\
n_1|_{z=-L_2}&=n_{1*}^{(0)}-\gamma_2\Theta\,.
\label{eq-08}
\end{align}

The condition of molecule flux balance on the liquid--vapour
interfaces is to be accounted as well. This condition yields the
vapour layer growth rate. The variation of the number of molecules
of one specie above the surface of the other specie liquid is
driven by diffusion only. Thus, if there is an increase of the
vapour layer thickness by $\delta{L_2}$ owned by evaporation from
the liquid 2 in time interval $\delta{t}$, the molecules of specie
1 can populate this added layer $\delta{L_2}$ only due to
diffusive influx from the bulk of the vapour layer in the same
time interval $\delta{t}$, which mathematically reads
\begin{align}
\left.n_1\right|_{z=-L_2}\delta L_{2}&=D_{12}\left.\frac{\partial n_{1}}
 {\partial z}\right|_{z=-L_2}\delta t\,,
\nonumber\\
\left.n_2\right|_{z=L_1}\delta L_{1}&=-D_{12}\left.\frac{\partial n_{2}}
 {\partial z}\right|_{z=L_1}\delta t\,.
\nonumber
\end{align}
These equalities yield
\begin{align}
\dot{L}_2&=\frac{D_{12}}{\left.n_1\right|_{z=-L_2}}
 \left.\frac{\partial n_1}{\partial z}\right|_{z=-L_2}
\approx \frac{D_{12}}{n_{1\ast}^{(0)}}
 \left.\frac{\partial n_1}{\partial z}\right|_{z=-L_2},
\label{eq-09}\\
\dot{L}_1&=-\frac{D_{12}}{\left.n_2\right|_{z=L_1}}
 \left.\frac{\partial n_2}{\partial z}\right|_{z=L_1}
\approx -\frac{D_{12}}{n_{2\ast}^{(0)}}
 \left.\frac{\partial n_2}{\partial z}\right|_{z=L_1},
\label{eq-10}
\end{align}
where the dot symbol denotes the time-derivative; for the
approximate equalities, we neglected relative corrections of order
 $\delta n_j^{(0)}/n_j^{(0)}$
which do not affect the leading order of accuracy. For the total
thickness $L=L_1+L_2$,
\begin{equation}
\dot{L}=D_{12}\left(\frac{1}{n_{1\ast}^{(0)}}
 \left.\frac{\partial n_1}{\partial z}\right|_{z=-L_2}
 -\frac{1}{n_{2\ast}^{(0)}}
 \left.\frac{\partial n_2}{\partial z}\right|_{z=L_1}\right).
\label{eq-11}
\end{equation}

It is convenient to introduce quantifiers for the liquid phase
evaporation rate. Let $v_{lj}$ be the velocity of the
$j$-th liquid in the reference frame fixed to its surface
($v_{l1}<0$, $v_{l2}>0$); in other terms, $|v_{lj}|$ is the speed
of the liquid phase retreat owned by evaporation. This velocity to
be calculated from the particle conservation condition as follows.
Let us consider the retreat of the surface of liquid 2 for
$v_{l2}\delta t$ owned by evaporation of the molecules into the
vapour layer. The number of molecules evaporated from the area
$S$, $n_{l2}\cdot S\cdot(v_{l2}\delta t)$ (where $n_{l2}$ is the
number density in the 2nd liquid phase) partially fills the
newly-formed ``slice'' $\delta L_2$ of the vapour layer with
particles of sort 2 and partially diffusively outflows from this
slice downhill the number density gradient, deeper into the vapour
layer;
$$
n_{l2} S v_{l2}\delta t=\left.n_2^{(0)}\right|_{z=-L_2}S\delta L_2
 -D_{12}\left.\frac{\partial n_2}{\partial z}\right|_{z=-L_2}
 S\delta t\,.
$$
Since $dn_2=-dn_1$ (see Eq.~(\ref{eq-01})), the number density
gradient at $z=-L_2$ can be taken from Eq.~(\ref{eq-09}) and the
latter equation yields (to the leading order of accuracy)
\begin{equation}
v_{l2}\approx\frac{n_{0*}}{n_{l2}}\dot{L}_2\,;
\label{eq-12}
\end{equation}
similarly,
\begin{equation}
v_{l1}\approx-\frac{n_{0*}}{n_{l1}}\dot{L}_1
\label{eq-13}
\end{equation}
(here, recall, $n_{lj}$ is the molecule number density in the
$j$-th liquid phase).

Let us now consider the energy balance in the system. The total
heat $Q_S$ supplied from the bulk of two liquids to the boiling
interface per its unit area,
\begin{equation}
\dot{Q}_S
 =\kappa_1\left.\frac{\partial T_1}{\partial z_1}\right|_{z_1=0}
 -\left.\kappa_2\frac{\partial T_2}{\partial z_2}\right|_{z_2=0}
\label{eq-14}
\end{equation}
(here $\kappa_i$ is the heat conductivity coefficient), is
consumed for evaporation from the liquid surfaces;
\begin{equation}
\dot{Q}_S = \Lambda_1n_{l1}(-v_{l1})+\Lambda_2n_{l2}v_{l2}\,,
\label{eq-15}
\end{equation}
where $\Lambda_j$ is the enthalpy of vaporization per one
molecule. In order to describe the difference between heat
influxes from the bulk of two liquids, one needs an additional
quantifier
$$
\dot{q}_S=\kappa_1\left.\frac{\partial T_1}{\partial z_1}\right|_{z_1=0}
 -\Bigg(-\left.\kappa_2\frac{\partial T_2}{\partial z_2}\right|_{z_2=0}\Bigg).
$$
Then, one can define the temperature boundary conditions for the
system;
\begin{align}
\left.\frac{\partial T_1}{\partial z_1}\right|_{z_1=0} &= +\frac{1}{2\kappa_1}\left(\dot{Q}_S+\dot{q}_S\right),
\label{eq-16}\\
\left.\frac{\partial T_2}{\partial z_2}\right|_{z_2=0} &= -\frac{1}{2\kappa_2}\left(\dot{Q}_S-\dot{q}_S\right).
\label{eq-17}
\end{align}

The heat conduction in the liquid phases is described by the
equation
\begin{equation}
\frac{\partial T_j}{\partial t}+v_{lj}\frac{\partial T_j}{\partial z_j}
 =\chi_j\frac{\partial^2T_j}{\partial z_j^2}\,,
\label{eq-18}
\end{equation}
where $\chi_j$ is the temperature diffusivity coefficient.

The evolution of the vapour layer is completely specified by
Eqs.~(\ref{eq-03}) and (\ref{eq-18}) with the boundary conditions
(\ref{eq-05})--(\ref{eq-08}) and (\ref{eq-16}), (\ref{eq-17}),
where $\dot L$ is given by Eq.~(\ref{eq-11}), $v_{lj}$ are given
by Eqs.~(\ref{eq-12}) and (\ref{eq-13}), $\dot{Q}_S$ is given by
Eq.~(\ref{eq-15}).

\section{Solution to the equations of the vapour layer evolution}
\label{sec:layer_growth}
According to the results derived in
Appendix~\ref{sec:app:distribution}, Eqs.~(\ref{eq-04}) with
boundary conditions~(\ref{eq-05})--(\ref{eq-08}) admit the
solution with nearly-linear concentration profiles:
\begin{align}
n_1(z,t)=n_{1\ast}^{(0)}-\gamma_2\Theta(t)+\alpha(z+L_2(t))
 +\mathcal{O}_1\left(\textstyle\frac{(\alpha L)^2}{n_{0\ast}}\right),
\label{eq-19}
 \\
n_2(z,t)=n_{2\ast}^{(0)}-\gamma_1\Theta(t)-\alpha(z-L_1(t))
 +\mathcal{O}_2\left(\textstyle\frac{(\alpha L)^2}{n_{0\ast}}\right),
\label{eq-20}
\end{align}
where
$$
\alpha=(\gamma_1+\gamma_2)\frac{\Theta(t)}{L(t)}
$$
is nearly constant in time, $\Theta(t)\propto t$, $L(t)\propto t$.
For these profiles, Eqs.~(\ref{eq-09})--(\ref{eq-11}) read
\begin{align}
\dot{L}_1&=\frac{D_{12}\alpha}{n_{2\ast}^{(0)}}\,,
\label{eq-21}
 \\
\dot{L}_2&=\frac{D_{12}\alpha}{n_{1\ast}^{(0)}}\,,
\label{eq-22}
 \\
L&=\frac{n_{0\ast}}
 {n_{1\ast}^{(0)}n_{2\ast}^{(0)}}D_{12}\alpha t\,,
\label{eq-23}
\end{align}
and Eq.~(\ref{eq-15}) yields
\begin{equation}
\dot{Q}_S=(\Lambda_1n_{1\ast}^{(0)}+\Lambda_2n_{2\ast}^{(0)})\dot{L}\,.
\label{eq-24}
\end{equation}

In the following discourse we will consider the second liquid, the
consideration for the first liquid can be constructed in the same
way. The behaviour of this subsystem is determined by
Eq.~(\ref{eq-18}) with boundary condition (\ref{eq-17}) and
$$
\left.\Theta_2\right|_{z=-L_2}=\Theta
 =\frac{\alpha L}{\gamma_1+\gamma_2}\,.
$$
This equation (\ref{eq-18}) with specified boundary conditions can
be solved in the same manner as for the symmetric case of two
liquids with similar
properties~\cite{Pimenova-Goldobin-JETP-2014}. Seeking the
solution in form
\begin{equation}
\nonumber
\Theta_2(z_2,t)=Ct+\theta(z_2)
\end{equation}
for $z_2\le0$, one can obtain
\begin{align}
 \Theta_2&=Ct+\frac{C}{v_{l2}}|z_2|
\nonumber\\
 &{}+\frac{\chi_2}{v_{l2}}
 \left(\frac{\dot{Q}_S-\dot{q}_S}{2\kappa_2}-\frac{C}{v_{l2}}\right)
 \left[1-\exp\left(-\frac{v_{l2}}{\chi_2}|z_2|\right)\right],
\label{eq-25}
\end{align}
where
$$
C=\frac{D_{12}}{\gamma_1+\gamma_2}\frac{n_{0\ast}}
 {n_{1\ast}^{(0)}n_{2\ast}^{(0)}}\alpha^2\,.
$$

Introduce new dimensionless variable
\begin{equation}
Z_2=\frac{v_{l2}}{\chi_2}|z_2|
 =\frac{n_{2*}^{(0)}}{n_{l2}}\frac{\dot{L}}{\chi_2}|z_2|\,.
 \label{eq-26}
\end{equation}
Then Eq.~(\ref{eq-25}) acquires the form
\begin{equation}
\Theta_2=Ct
 +\frac{\chi_2}{v_{l2}}\frac{\dot{Q}_S-\dot{q}_S}{2\kappa_2}(1-e^{-Z_2})
 +C\frac{\chi_2}{v_{l2}^2}(Z_2-1+e^{-Z_2})\,.
\label{eq-27}
\end{equation}
With substitution of typical parameter values it can be shown (see
Ref.~\cite{Pimenova-Goldobin-JETP-2014} as well), that the
characteristic values of the amplitudes of the second and third
terms in the last equation are huge; for water $10^2\,\mathrm{K}$
and $10^5\,\mathrm{K}$, respectively. Since $\Theta_2$ is not
larger than $10$--$20\,\mathrm{K}$ even in extreme cases, the
argument $Z_2$ of the second and third terms should be small.
However, one cannot plainly use the linear in $Z_2$ approximation,
since the expansion of the third term has vanishing linear part
and starts from $Z_2^2$, while its amplitude is by three orders of
magnitude larger than that of the second term. Hence, we keep
leading contributions from the both terms: linear in $Z_2$ for the
second term and quadratic in $Z_2$ for the third term. Simplified
equation~(\ref{eq-27}) reads
\begin{equation}
\Theta_2\approx
 Ct+\frac{\chi_2}{v_{l2}}\frac{\dot{Q}_S-\dot{q}_S}{2\kappa_2}Z_2
 +\frac{C}{2}\frac{\chi_2}{v_{l2}^2}Z_2^2\,.
\label{eq-28}
\end{equation}
The spatial part of the field $\Theta$ has a structure
$\theta(z)\sim10^2\mathrm{K}\times Z_2+10^5\mathrm{K}\times Z_2^2$.
The parameter $Z_2$ is rather small. It is close to $10^{-2}$ even
for the maximal possible overheating
$\theta(z)\sim10\,\mathrm{K}$; for a stronger overheating the bulk
boiling in one of liquids occurs. For the overheating smaller than
$0.1\,\mathrm{K}$ the main contribution in $\theta(z)$ is made by
the linear term, while for larger overheating the quadratic term
dominates.

In more natural terms of physical characteristics $\dot{L}$ and
$\dot{Q}_S$, Eq.~(\ref{eq-28}) reads
\begin{align}
\nonumber
\Theta_2=&\frac{n_{1\ast}^{(0)}n_{2\ast}^{(0)}\,\dot{L}^2}{(\gamma_1+\gamma_2)D_{12}n_{0\ast}}t
 +\frac{\Lambda_1n_{1*}^{(0)}+\Lambda_2n_{2*}^{(0)}}{2n_{2*}^{(0)}c_{p,l2}}
 \frac{\dot{Q}_S-\dot{q}_S}{\dot{Q}_S}Z_2\\
&{}+\frac{1}{2}\frac{n_{l2}^2\,n_{1\ast}^{(0)}}
 {(\gamma_1+\gamma_2)\,n_{0\ast}\,n_{2\ast}^{(0)}}
 \frac{\chi_2}{D_{12}}Z_2^2\,,
\label{eq-29}
\end{align}
where $c_{p,lj}$ is the specific heat per one molecule in $j$-th
liquid under the constant pressure conditions. For identical
physical properties of two liquids and no heat influx asymmetry
$\dot{q}_S$, the last equation takes the form of Eq.~(29)
in~\cite{Pimenova-Goldobin-JETP-2014}, derived for an idealised
symmetric case.

Similarly, for the first liquid one can find
\begin{align}
\nonumber
\Theta_1=&\frac{n_{1\ast}^{(0)}n_{2\ast}^{(0)}\,\dot{L}^2}{(\gamma_1+\gamma_2)D_{12}n_{0\ast}}t
 +\frac{\Lambda_1n_{1*}^{(0)}+\Lambda_2n_{2*}^{(0)}}{2n_{1*}^{(0)}c_{p,l1}}
 \frac{\dot{Q}_S+\dot{q}_S}{\dot{Q}_S}Z_1\\
&{}+\frac{1}{2}\frac{n_{l1}^2\,n_{2\ast}^{(0)}}
 {(\gamma_1+\gamma_2)\,n_{0\ast}\,n_{1\ast}^{(0)}}
 \frac{\chi_1}{D_{12}}Z_1^2\,.
\label{eq-30}
\end{align}

All material parameters appearing in Eqs.~(\ref{eq-24}),
(\ref{eq-29}), and (\ref{eq-30}) are provided in Tab.~\ref{tab1}.
For calculations of the interfacial boiling point $T_\ast$ and
physical properties of vapour mixture at $T_\ast$ see
Appendix~\ref{sec:app:params}.

\begin{table}[t]
\caption{Chemical physical properties of water and n-heptane at their interface boiling temperature $T_\ast$ and $P_0=1\,\mathrm{atm}$.}
\begin{center}
\begin{tabular}{|c|c|c|}
\hline
 & $\mathrm{H_2O}$ & \mbox{n-heptane}
 \\
\hline
\vspace{-7pt}
&\qquad\qquad\qquad\quad&\qquad\qquad\qquad\quad
 \\
Bulk boiling point (K)
 & $373.15$ & $371.58$ \\[5pt]
$T_\ast$ (K)
 & \multicolumn{2}{|c|}{$351.71$ ($=78.56^\circ\mathrm{C}$)} \\[5pt]
$\Lambda_j/k_\mathrm{B}$ (K)
 & $4987$ & $3977$ \\[5pt]
$c_{p,lj}/k_\mathrm{B}$
 & $9.09$ & $27.02$ \\[5pt]
$\chi_j$ (m$^2$/s)
 & $1.70\cdot10^{-7}$ & $0.66\cdot10^{-7}$ \\[5pt]
$\rho_{lj}$ (kg/m$^3$)
 & $0.973\cdot10^3$ & $0.638\cdot10^3$ \\[5pt]
$n_{lj}/n_{0\ast}$
 & $1.559\cdot10^3$ & $0.184\cdot10^3$ \\[5pt]
$n_{j\ast}^{(0)}/n_{0\ast}$
 & $0.446$ & $0.554$ \\[5pt]
$\gamma_j/n_{0\ast}$ (K$^{-1}$)
 & $0.0180$ & $0.0177$ \\[5pt]
$D_{12}(T_\ast)$ (m$^2$/s)
 & \multicolumn{2}{|c|}{$1.20\cdot10^{-5}$} \\[5pt]
$\eta_{12}(T_\ast)$ (Pa$\cdot$s)
 & \multicolumn{2}{|c|}{$0.59\cdot10^{-5}$} \\[5pt]
$\sigma_{j}$ (N/m)
 & $62.93\cdot10^{-3}$ & $14.40\cdot10^{-3}$ \\[3pt]
\hline
\end{tabular}
\end{center}
\label{tab1}
\end{table}

The mathematical model developed in Sec.~\ref{sec:model} and the
rigorous solution for the vapour layer growth, derived in this
section, belong to our main findings we report with this paper. In
the following sections we treat relationships between the solution
we derived and macroscopic characteristics of the state of the
system experiencing interfacial boiling.

\section{Relationships between kinetics of the vapour layer and mean macroscopic parameters of the system}
\label{sec:macroscopic}
In this section we perform assessments of the characteristics of
the steady process of boiling. A statistically stationary regime
can be described in terms of mean heat influx, mean overheating
degree and evaporation rate, where the latter two are maintained
by the former. Two cases are addressed here: (i)~a system well
stirred by boiling and (ii)~a stratified system (as in
Figs.~\ref{fig1}(b), \ref{fig2}(a)). First, we consider the
limitation on the growth of the vapour layer due to its buoyancy,
which leads to formation of vapour bubbles and breakaway of the
layer. Then, on the basis of the results of this consideration, we
evaluate the state of the system with given heat inflow.
Note, while the results of previous sections are rigorous, our
considerations in this section is approximate; our main task here
is to develop a qualitative description of the macroscopic system
behaviour and to interpret the analytical solutions derived.

\subsection{Breakaway of vapour layer}
\label{sec:breakaway}

\subsubsection{Case of well-stirred system}
In the case where the system is considered to be an emulsion of
two liquids, well-mixed by the process of boiling, a significant
parameter of the system state is the mean interface area per unit
volume, $\delta{S}/\delta{V}$. This value depends on parameters of
liquids and characteristics of the evaporation process, which are
controlled by the mean overheating and the bubble production
rate~\cite{Filipczak-etal-2011}. In this work the volumes of both
components are assumed to be commensurable, no phase can be
considered as a medium hosting inclusions of the other phase. The
characteristic width of the neighborhood of the vapour layer,
beyond which the neighborhood of another vapour layer lies, is
\begin{equation}
\nonumber
H_1+H_2\sim\left(\frac{\delta{S}}{\delta{V}}\right)^{-1}.
\end{equation}
The relationship between characteristic thicknesses of the liquid
layers $H_1$ and $H_2$ is
$$
\frac{H_1}{H_2}=\frac{\phi_1}{\phi_2}=\frac{\phi_1}{1-\phi_1}\,,
$$
where $\phi_j$ is the volumetric fraction of the $j$-th liquid in
the system. It will be convenient to use
\begin{equation}
H_j\sim\phi_j\left(\frac{\delta{S}}{\delta{V}}\right)^{-1}.
\label{eq-31}
\end{equation}

The process of boiling of a mixture above the bulk boiling
temperature of the more volatile liquid is well-addressed in the
literature\cite{Simpson-etal-1974,Celata-1995,Roesle-Kulacki-2012-1,Roesle-Kulacki-2012-2,Sideman-Isenberg-1967,Kendoush-2004,Filipczak-etal-2011}.
Hydrodynamic aspects of the process of boiling below the bulk
boiling temperature has to be essentially similar at the
macroscopic level; rising vapour bubbles drives the stirring of
system, working against the gravitational stratification into two
layers with a flat horizontal interface, the surface tension
forces tending to minimize the interface area, and viscous
dissipation of the flow kinetic energy. Specifically, the
behaviour of parameter $(\delta{S}/\delta{V})$ depending on
macroscopic characteristics of processes in the system should be
the same as for systems with superheating of the more volatile
component. In Appendix~\ref{sec:app:dSdV} we additionally provide
an analytical assessment of the dependence of
$(\delta{S}/\delta{V})$ on the evaporation rate (or heat influx)
for a well-stirred system.

\begin{figure}[t]
\center{
\includegraphics[width=0.45\textwidth]%
{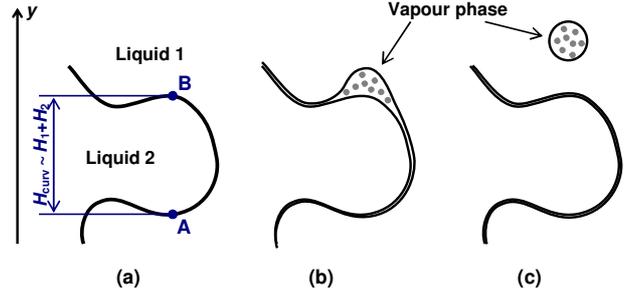}}
\caption{
Process of bubble formation from vapour layer in a well-stirred
system.
 }
\label{fig4}
\end{figure}

The growth of the vapour layer is limited by its buoyancy; when
the layer becomes too thick, vapour driven upwards by the pressure
gradient can seepage along the layer quite efficiently and forms
bubbles (see Fig.~\ref{fig4}). Separating from the layer these
bubbles entrain vapour from it and thus effectively reset the
layer to the zero-thickness state. In
Ref.~\cite{Pimenova-Goldobin-JETP-2014} this process was
considered in detail for the symmetric case and the consideration
can be plainly repeated for the non-symmetric case we consider
here. With this consideration one can find the characteristic
contribution of the Poiseuille's viscous seepage of vapour along a
thin layer into the layer thickness derivative $\dot{L}$:
$$
\dot{L}_p\approx-\frac{(\rho_{l1}+\rho_{l2})gL^3}{96\,\eta_{12}}
 \left(\frac{\delta{S}}{\delta{V}}\right)\,,
$$
where $\eta_{12}$ is the dynamic viscosity of vapour, $g$ is the
gravity, $\rho_j$ are the densities of liquids.

One can notice, that $\dot{L}_p$ strongly depends on the layer
thickness, $\dot{L}_p\propto L^3$, as compared to the growth owned
by evaporation, which has a constant rate $\dot{L}$. For this
reason, one can neglect the role of the vapour seepage at early
stage and consider that the vapour layer breaks away when
$\dot{L}_p$ becomes equal to the evaporational growth rate. Hence,
the vapour layer thickness attained before the breakaway is
$$
L_\ast\approx\left[\frac{96\,\eta_{12}}{(\rho_{l1}+\rho_{l2})g}
 \left(\frac{\delta{S}}{\delta{V}}\right)^{-1}\dot{L}\right]^{1/3}\,,
$$
which corresponds to the time instant
\begin{equation}
t_\ast\approx\frac{L_\ast}{\dot{L}}
\approx\left[\frac{96\,\eta_{12}}{(\rho_{l1}+\rho_{l2})g}
\left(\frac{\delta{S}}{\delta{V}}\right)^{-1}\right]^{1/3}\dot{L}^{-2/3}.
\label{eq-32}
\end{equation}

It should be noted that when the vapour layer breaks away, some
overheating of liquid, related to the linear in time term in
Eqs.~(\ref{eq-29})--(\ref{eq-30}), remains in the vicinity of the
layer. However, for the symmetric case, the heat of this
overheating was revealed to make a negligible contribution into
the net heat balance~\cite{Pimenova-Goldobin-JETP-2014}. For the
asymmetric case the orders of magnitude of values are the same and
this overheat can be also neglected; suggesting that after
breakaway the interface state is reset to the state
(\ref{eq-29})--(\ref{eq-30}) with $t=0$ and $L=0$.

With the known reference time instant $t_\ast$ of the vapour layer
resetting, one can evaluate the characteristic maximal temperature
in components attained at the maximal distance from the interface,
$z_{j,\mathrm{max}}\sim H_j$, at $t=t_\ast$. Eq.~(\ref{eq-29})
yields
\begin{align}
\Theta_{2,\mathrm{max}}=&\frac{n_{1\ast}^{(0)}n_{2\ast}^{(0)}\,\dot{L}^2}
 {(\gamma_1+\gamma_2)D_{12}n_{0\ast}}t_\ast
 \nonumber\\
&{}+\frac{\Lambda_1n_{1*}^{(0)}+\Lambda_2n_{2*}^{(0)}}{2n_{2*}^{(0)}c_{p,l2}}
 \left[1-\frac{\dot{q}_S}{\dot{Q}_S}\right]Z_{2,\mathrm{max}}
 \nonumber\\
&{}+\frac{1}{2}\frac{n_{l2}^2\,n_{1\ast}^{(0)}}
 {(\gamma_1+\gamma_2)\,n_{0\ast}\,n_{2\ast}^{(0)}}
 \frac{\chi_2}{D_{12}}Z_{2,\mathrm{max}}^2
\label{eq-33}
\\
 &\hspace{-10mm}=\Theta_{4/3}Z_\mathrm{max}^{4/3}
 +\left[1-\frac{\dot{q}_S}{\dot{Q}_S}\right]\Theta_{2,1}Z_\mathrm{max}
 +\Theta_{2,2}Z_\mathrm{max}^2\,,
\label{eq-34}
\end{align}
where, in accordance to Eqs.~(\ref{eq-32}), (\ref{eq-26}) and (\ref{eq-31}),
\begin{align}
Z_\mathrm{max}&=Z_{1,\mathrm{max}}+Z_{2,\mathrm{max}}
\nonumber\\[5pt]
 &=\left(\frac{n_{1*}^{(0)}}{n_{l1}}\frac{\phi_1}{\chi_1}
 +\frac{n_{2*}^{(0)}}{n_{l2}}\frac{\phi_2}{\chi_2}\right)
 \frac{\dot{L}}{(\delta{S}/\delta{V})}
\label{eq-35}
\end{align}
is introduced so that
$$
\displaystyle
Z_{j,\mathrm{max}}=\psi_jZ_\mathrm{max}\,,\qquad
\psi_j=\frac{\frac{n_{j*}^{(0)}}{n_{lj}}\frac{\phi_j}{\chi_j}}
 {\frac{n_{1*}^{(0)}}{n_{l1}}\frac{\phi_1}{\chi_1}
 +\frac{n_{2*}^{(0)}}{n_{l2}}\frac{\phi_2}{\chi_2}}\,,
$$
and
\begin{equation}
\Theta_{4/3}=\frac{\displaystyle
 \frac{n_{1\ast}^{(0)}n_{2\ast}^{(0)}}{(\gamma_1+\gamma_2)D_{12}n_{0\ast}}
 \left(\frac{96\,\eta_{12}}{(\rho_1+\rho_2)g}\right)^{1/3}
 \frac{\delta{S}}{\delta{V}}}
 {\displaystyle
 \left(\frac{n_{1*}^{(0)}}{n_{l1}}\frac{\phi_1}{\chi_1}
 +\frac{n_{2*}^{(0)}}{n_{l2}}\frac{\phi_2}{\chi_2}\right)^{4/3}}\,,
\label{eq-36}
\end{equation}
\begin{equation}
\Theta_{j,1}=\frac{\Lambda_1n_{1*}^{(0)}+\Lambda_2n_{2*}^{(0)}}
 {2n_{j*}^{(0)}c_{p,lj}}\,\psi_j\,,
\label{eq-37}
\end{equation}
\begin{equation}
\Theta_{j,2}=\frac{1}{2}\frac{n_{lj}^2\,n_{2-j\,\ast}^{(0)}}
 {(\gamma_1+\gamma_2)\,n_{0\ast}\,n_{j\ast}^{(0)}}
 \frac{\chi_j}{D_{12}}\,\psi_j^2\,.
\label{eq-38}
\end{equation}
Similarly to Eq.~(\ref{eq-34}), the characteristic maximal
temperature of the component 1 can be written down;
\begin{align}
\Theta_{1,\mathrm{max}}=\Theta_{4/3}Z_\mathrm{max}^{4/3}
 +\left[1+\frac{\dot{q}_S}{\dot{Q}_S}\right]\Theta_{1,1}Z_\mathrm{max}
 +\Theta_{1,2}Z_\mathrm{max}^2\,.
\label{eq-39}
\end{align}

As discussed above, after breakaway of the vapour layer, one can
approximately assume the interface and its vicinity to be reset to
the early stage of the vapour-layer-growth solution, when $L\ll
L_\ast$. Then the average over time and space values of terms in
Eqs.~(\ref{eq-29})--(\ref{eq-30}) are determined by the averages
$\langle{t}\rangle=t_\ast/2$ and
$\langle{z^n}\rangle=z_\mathrm{max}^n/(n+1)$;
\begin{align}
\langle\Theta_1\rangle=&\frac{\Theta_{4/3}}{2}Z_\mathrm{max}^{4/3}
 +\left[1+\frac{\dot{q}_S}{\dot{Q}_S}\right]\frac{\Theta_{1,1}}{2}Z_\mathrm{max}
 +\frac{\Theta_{1,2}}{3}Z_\mathrm{max}^2\,,
\label{eq-40}
\\[5pt]
\langle\Theta_2\rangle=&\frac{\Theta_{4/3}}{2}Z_\mathrm{max}^{4/3}
 +\left[1-\frac{\dot{q}_S}{\dot{Q}_S}\right]\frac{\Theta_{2,1}}{2}Z_\mathrm{max}
 +\frac{\Theta_{2,2}}{3}Z_\mathrm{max}^2\,.
\label{eq-41}
\end{align}

Eqs.~(\ref{eq-34}), (\ref{eq-39}), (\ref{eq-40}), and
(\ref{eq-41}) provide relations between the system state variables
and the variables $\dot{L}$ (or $Z_\mathrm{max}$, see
Eq.~(\ref{eq-35})) and heat inflow asymmetry
$(\dot{q}_S/\dot{Q}_S)$. While the latter two are not accessible
for direct control, the former can be manipulated directly. Given
$\Theta_{j,\mathrm{max}}$ are maintained to be fixed,
$(\dot{q}_S/\dot{Q}_S)$ and $Z_\mathrm{max}$ can be calculated
from Eqs.~(\ref{eq-34}) and (\ref{eq-39}); the equation
\begin{align}
\frac{\Theta_{1,\mathrm{max}}}{\Theta_{1,1}}
 +\frac{\Theta_{2,\mathrm{max}}}{\Theta_{2,1}}
 =\left(\frac{1}{\Theta_{1,1}}+\frac{1}{\Theta_{2,1}}\right)
 \Theta_{4/3}Z_\mathrm{max}^{4/3}\quad
\nonumber\\
 {}+2Z_\mathrm{max}
 +\left(\frac{\Theta_{1,2}}{\Theta_{1,1}}
 +\frac{\Theta_{2,2}}{\Theta_{2,1}}\right)Z_\mathrm{max}^2
\label{eq-42}
\end{align}
governs $Z_\mathrm{max}$ and, with calculated $Z_\mathrm{max}$,
one can straightforwardly find $(\dot{q}_S/\dot{Q}_S)$ from
Eq.~(\ref{eq-34}) or (\ref{eq-39}). Similarly, given
$\langle\Theta_j\rangle$ are maintained to be fixed,
\begin{align}
\frac{\langle\Theta_1\rangle}{\Theta_{1,1}}
 +\frac{\langle\Theta_2\rangle}{\Theta_{2,1}}
 =\left(\frac{1}{\Theta_{1,1}}+\frac{1}{\Theta_{2,1}}\right)
 \frac{\Theta_{4/3}}{2}Z_\mathrm{max}^{4/3}\qquad
\nonumber\\
 {}+Z_\mathrm{max}
 +\frac{1}{3}\left(\frac{\Theta_{1,2}}{\Theta_{1,1}}
 +\frac{\Theta_{2,2}}{\Theta_{2,1}}\right)Z_\mathrm{max}^2\,.
\label{eq-43}
\end{align}

\subsubsection{Case of stratified system}
The process of boiling can be not strong enough for the rising
vapour bubbles to enforce any significant stirring of the system.
For instance, one can observe such a behaviour of the system in
Figs.~\ref{fig1}(b) and~\ref{fig2}(a). In this case system is well
stratified; the light liquid rests upon the heavy one with mainly
unperturbed interface. The breakaway of the vapour layer in such a
system is related to the Rayleigh--Taylor
instability~\cite{Rayleigh-1883,Taylor-1950} of the upper
vapour--water interface (one can see Fig.~\ref{fig5} in
Appendix~\ref{sec:app:instability}), which is gravitationally
unstable.

However, our case is significantly different compared to the
conventional Rayleigh--Taylor instability; we deal with an
extremely thin vapour layer in between of two liquids, without
which the system is stably stratified. This specific case of
Rayleigh--Taylor instability is actualised by our problem setup
and, to the authors' knowledge, was not addressed in the
literature; the consideration of this instability is provided in
Appendix~\ref{sec:app:instability}. Without vapour generation, the
exponential growth rate of the most dangerous perturbations is
accurately given by Eq.~(\ref{eq-apc-25});
\begin{equation}
\lambda_1=\frac{L^3}{54\eta_{12}}\frac{\sigma_1\sigma_2}{\sigma_1+\sigma_2}
 \frac{k_1^4(k_2^2+k_1^2/3)}{k_{12}^2+k_1^2/3}\,,
\label{eq-44}
\end{equation}
where $\sigma_j$ are the surface tension coefficients for
vapour--liquid interfaces and $k_1$, $k_2$, and $k_{12}$ are given
by Eqs.~(\ref{eq-apc-21}). To be able to track the physical
meaning of terms in the latter equation, we introduce
$\tilde\sigma=\sigma_1\sigma_2/(\sigma_1+\sigma_2)$ and
$\tilde\rho$ in a way that
 $(\tilde\rho{g}/\tilde\sigma)^2=k_1^4(k_2^2+k_1^2/3)/(k_{12}^2+k_1^2/3)$
(compare to Eqs.~(\ref{eq-apc-21})). Then, Eq.~(\ref{eq-44}) reads
$$
\lambda_1=\frac{L^3(\tilde\rho g)^2}{54\eta_{12}\tilde\sigma}\,.
$$

The reference time of hydrodynamic instability development
$t_\ast\sim1/\lambda_1\propto L^{-3}$. Again, one can notice the
instability development to be very slow for small $L$ and
extremely fast for large $L$. In the same spirit as for the
previous case, we assume the vapour layer to growth with
negligible effect of the instability until the instability
development time $t_\ast$ becomes commensurable to the layer
growth time $L/\dot{L}$ and fast layer breakaway happens. Thus,
$$
t_\ast=\frac{54\eta_{12}\tilde\sigma}
 {(\dot{L}t_\ast)^3(\tilde\rho g)^2}\,,
$$
and one finds
\begin{equation}
t_\ast=\frac{(54\eta_{12}\tilde\sigma)^{1/4}}
 {(\tilde\rho g)^{1/2}}\dot{L}^{-3/4}\,.
\label{eq-45}
\end{equation}
Noteworthy, this case is featured by a different power law of
dependence $t_\ast(\dot{L})$ than in Eq.~(\ref{eq-32}).

With the reference time $t_\ast$ of the vapour layer resetting
given by Eq.~(\ref{eq-45}), one can evaluate the characteristic
maximal temperature in components similarly to the case of a
well-stirred system. Eqs.~(\ref{eq-29}) and (\ref{eq-30}) yield
for a stratified system
\begin{align}
\Theta_{1,\mathrm{max}}=&\Theta_{5/4}Z_\mathrm{max}^{5/4}
 +\left[1+\frac{\dot{q}_S}{\dot{Q}_S}\right]\Theta_{1,1}Z_\mathrm{max}
 +\Theta_{1,2}Z_\mathrm{max}^2\,,
\label{eq-46}
\\
\Theta_{2,\mathrm{max}}=&\Theta_{5/4}Z_\mathrm{max}^{5/4}
 +\left[1-\frac{\dot{q}_S}{\dot{Q}_S}\right]\Theta_{2,1}Z_\mathrm{max}
 +\Theta_{2,2}Z_\mathrm{max}^2\,,
\label{eq-47}
\end{align}
where
\begin{equation}
\Theta_{5/4}=\frac{\displaystyle
 \frac{n_{1\ast}^{(0)}n_{2\ast}^{(0)}}{(\gamma_1+\gamma_2)D_{12}n_{0\ast}}
 \frac{(54\eta_{12}\tilde{\sigma})^{1/4}}{(\tilde\rho g)^{1/2}}
 \left(\frac{\delta{S}}{\delta{V}}\right)^{5/4}}
 {\displaystyle
 \left(\frac{n_{1*}^{(0)}}{n_{l1}}\frac{\phi_1}{\chi_1}
 +\frac{n_{2*}^{(0)}}{n_{l2}}\frac{\phi_2}{\chi_2}\right)^{5/4}}\,.
\label{eq-48}
\end{equation}

Note, for this case it is more suitable to express volumetric
fractions $\phi_j$ of components in terms of well determined
parameters $H_j$, which are the thicknesses of two liquid layers;
$\phi_j=H_j/(H_1+H_2)$.

Mean temperatures are
\begin{align}
\langle\Theta_1\rangle=&\frac{\Theta_{5/4}}{2}Z_\mathrm{max}^{5/4}
 +\left[1+\frac{\dot{q}_S}{\dot{Q}_S}\right]\frac{\Theta_{1,1}}{2}Z_\mathrm{max}
 +\frac{\Theta_{1,2}}{3}Z_\mathrm{max}^2\,,
\label{eq-49}
\\[5pt]
\langle\Theta_2\rangle=&\frac{\Theta_{5/4}}{2}Z_\mathrm{max}^{5/4}
 +\left[1-\frac{\dot{q}_S}{\dot{Q}_S}\right]\frac{\Theta_{2,1}}{2}Z_\mathrm{max}
 +\frac{\Theta_{2,2}}{3}Z_\mathrm{max}^2\,.
\label{eq-50}
\end{align}

For fixed $\Theta_{j,\mathrm{max}}$ or $\langle\Theta_j\rangle$, one finds
\begin{align}
\frac{\Theta_{1,\mathrm{max}}}{\Theta_{1,1}}
 +\frac{\Theta_{2,\mathrm{max}}}{\Theta_{2,1}}
 =\left(\frac{1}{\Theta_{1,1}}+\frac{1}{\Theta_{2,1}}\right)
 \Theta_{5/4}Z_\mathrm{max}^{5/4}\quad
\nonumber\\
 {}+2Z_\mathrm{max}
 +\left(\frac{\Theta_{1,2}}{\Theta_{1,1}}
 +\frac{\Theta_{2,2}}{\Theta_{2,1}}\right)Z_\mathrm{max}^2\,,
\label{eq-51}
\end{align}
\begin{align}
\frac{\langle\Theta_1\rangle}{\Theta_{1,1}}
 +\frac{\langle\Theta_2\rangle}{\Theta_{2,1}}
 =\left(\frac{1}{\Theta_{1,1}}+\frac{1}{\Theta_{2,1}}\right)
 \frac{\Theta_{5/4}}{2}Z_\mathrm{max}^{5/4}\qquad
\nonumber\\
 {}+Z_\mathrm{max}
 +\frac{1}{3}\left(\frac{\Theta_{1,2}}{\Theta_{1,1}}
 +\frac{\Theta_{2,2}}{\Theta_{2,1}}\right)Z_\mathrm{max}^2\,.
\label{eq-52}
\end{align}

\subsection{Vapour generation at constant heat inflow}
\label{sec:evaporation}
Let us establish the relation between the system state parameters
and the volumetric heat influx
$$
\dot{Q}_V=\frac{\delta Q}{\delta V \delta t}\,.
$$
Corresponding heat influx per unit area of the interface
$$
\dot{Q}_S=\left(\frac{\delta{S}}{\delta{V}}\right)^{-1}\dot{Q}_V.
$$
For a statistically stationary process of boiling, mean
temperature does not grow and all the heat influx to the system is
spent for vapour generation; therefore, Eq.~(\ref{eq-24}) for the
relation between $\dot{Q}_S$ and $\dot{L}$ is valid. Hence,
Eq.~(\ref{eq-35}) reads
\begin{align}
Z_\mathrm{max}&=\left(\frac{n_{1*}^{(0)}}{n_{l1}}\frac{\phi_1}{\chi_1}
 +\frac{n_{2*}^{(0)}}{n_{l2}}\frac{\phi_2}{\chi_2}\right)
 \frac{\dot{L}}{(\delta{S}/\delta{V})}
\nonumber\\[5pt]
 &=\frac{\displaystyle
 \left(\frac{n_{1*}^{(0)}}{n_{l1}}\frac{\phi_1}{\chi_1}
 +\frac{n_{2*}^{(0)}}{n_{l2}}\frac{\phi_2}{\chi_2}\right)
 \frac{\dot{Q}_V}{(\delta{S}/\delta{V})^2}}
 {\Lambda_1n_{1\ast}^{(0)}+\Lambda_2n_{2\ast}^{(0)}}\,.
\label{eq-53}
\end{align}
With this expression for $Z_\mathrm{max}$, one can calculate
maximal temperatures in components with Eqs.~(\ref{eq-39}),
(\ref{eq-33}) or (\ref{eq-46}), (\ref{eq-47}) and mean
temperatures with Eqs.~(\ref{eq-40}), (\ref{eq-41}) or
(\ref{eq-49}), (\ref{eq-50}) for the cases of well-stirred and
stratified systems.

\paragraph{Assessments: Combustion of n-heptane over water.}
For combustion of a light flammable liquid over a heavier liquid
one can evaluate the conductive heat influx from the burning
surface to the interface;
$\dot{Q}_S\sim\kappa_1(T_{1b}-T_\ast)/H_1$, where $T_{1b}$ is the
bulk boiling temperature of the burning liquid (see
Sec.~\ref{sec:example} for explanations why temperature of the
surface of the burning liquid must be nearly $T_{1b}$) and $H_1$
is the flammable liquid layer thickness. Here, for an estimate, we
neglect the heat conduction loss from liquids to the environment.
Eq.~(\ref{eq-24}) yields
\begin{align}
\dot{L}&=\frac{\kappa_1}{\Lambda_1n_{1\ast}^{(0)}+\Lambda_2n_{2\ast}^{(0)}}
 \frac{T_{1b}-T_\ast}{H_1}
\nonumber\\[5pt]
 & =\frac{c_{p,l1}n_{l1}\chi_1}{\Lambda_1n_{1\ast}^{(0)}+\Lambda_2n_{2\ast}^{(0)}}
 \frac{T_{1b}-T_\ast}{H_1}
 \approx\frac{1.47\cdot10^{-6}\mathrm{m^2/s}}{H_1}\,,
\nonumber
\end{align}
where parameter values for n-heptane--water are taken from
Tab.~\ref{tab1}. With Eq.~(\ref{eq-45}), one can calculate
$t_\ast\approx5.53\cdot10^{-4}\mathrm{m^{3/4}s^{1/4}}\times\dot{L}^{-3/4}$
and find the reference layer thickness
$L_\ast=\dot{L}t_\ast\approx5.53\cdot10^{-4}\mathrm{m^{3/4}s^{1/4}}\times\dot{L}^{1/4}
\approx1.93\cdot10^{-5}\mathrm{m^{5/4}}\times H_1^{-1/4}$ at the
instant of breakaway. The wavelength of the most dangerous
instability mode of a thin vapour layer between the n-heptane and
water layers
$l_{vl}=2\pi/k_{\mathrm{max}}^{n\mathrm{C_7H_{16}-H_2O}}\approx1.46\,\mathrm{cm}$
(see Appendix~\ref{sec:app:instability}). Hence, a vapour bubble
separating from the interface is formed from the vapour layer
patch of area $S_{vl}\approx l_{vl}^2$ and possesses volume
$V_b\approx S_{vl}L_\ast$ or $(\pi/6)d_b^3$, where $d_b$ is the
bubble diameter. Finally, the bubble diameter
$$
d_b\approx[(6/\pi)l_{vl}^2L_\ast]^{1/3}
 \approx2.0\cdot10^{-3}\mathrm{m^{13/12}}\times H_1^{-1/12}\,.
$$
One can notice the dependence of $d_b$ on $H_1$ to be extremely
``slow'', power $(-1/12)$; for the layer thickness of order of
magnitude of $1\,\mathrm{cm}$ the bubble diameter
$d_b\approx2.9\,\mathrm{mm}$, which slightly overestimates the
characteristic size observed in Fig.~\ref{fig2}(b). This
overestimation is expected because we neglected the heat loss to
the environment and thus overestimated the heat influx spent for
the generation of vapour. Thus our theoretical description of the
process yields results which match experimental observations well.

\section{Discussion of simplification assumptions}
\label{sec:simplific}
Let us summarise the simplification assumptions made in the course
of developing the theory and discuss possible inaccuracies brought
in with these assumptions.

\vspace{5pt}
\noindent$\bullet$\ The solution for the transversal structure of the
growing vapour layer and its vicinity is derived neglecting the
layer curvature and sideway motion of vapour and liquid. The
inaccuracy brought in with these neglections is expected to be
small by virtue of the smallness of the maximal vapour layer
thickness attained before the layer breaks away, which is
$L_\mathrm{max}\sim10^{-5}-10^{-4}\,\mathrm{m}$, compared to the
shortest scale along the layer, which is
$\sim10^{-3}-10^{-2}\,\mathrm{m}$.

\vspace{5pt}
\noindent$\bullet$\ For the evaporation process we neglect
finiteness of the rate of molecule escape from liquid into vapour,
assuming the vapour number density immediately above the liquid
surface to be equal the saturation vapour one. We provide
arguments for this assumption. In the light of final results this
assumption can be treated to work well enough as the predicted
features of the bubble formation process agree with experimental
observations, while for evaporation from open liquid surfaces the
finiteness of the escape rate leads to the decrease of the
evaporation rate by factor of $10$ (e.g.,
see~\cite{Anisimov-Rakhmatulina-1973}).

\vspace{5pt}
\noindent$\bullet$\ Considering the hydrodynamic instability of
the stratified three-layer system, we assume the liquid flow rates
to be small compared to the rates of the vapour flow along the
vapour layer (see Appendix~\ref{sec:app:instability}). The
accuracy of this approximation can be quantified by the ratio of
dynamic viscosities of vapour and liquid,
$\eta_{12}/\eta_\mathrm{liq}\approx10^{-2}$. The liquid flow is
also considered to be inviscid. Indeed, for the instability flow
in liquid the characteristic rate $v\sim
L_\mathrm{max}\lambda_{1,\mathrm{max}}\sim10^{-3}\,\mathrm{m/s}$
and the corresponding thickness of the viscous boundary layer
$h_\mathrm{v.b.l.}\sim\nu_\mathrm{liq}/v\sim1\,\mathrm{mm}$ is
small compared to the characteristic spatial scale of the
instability, which is $1.0-1.5\,\mathrm{cm}$.

\vspace{5pt}
\noindent$\bullet$\ The reference time of the layer breakaway is
calculated for two limit cases: a well-stirred system and a
well-stratified one. With estimations of
Appendix~\ref{sec:app:dSdV}, one can see that the boiling regime
is controlled by the heat inflow rate into the system, material
parameters and the system volume. For the conditions of the
demonstration experiment with n-heptane--water system
(Fig.\,\ref{fig2}), the system was observed to be rather close to
a well-stratified state. For the burning ``white spirit''--water
system (Fig.\,\ref{fig1}), we observed all the range of boiling
regimes from the one with prominent stratification to a strong
stirring.

\section{Conclusion}
\label{sec:conclusion}
We have theoretically explored the process of boiling at the
interface between two immiscible liquids below the bulk boiling
temperatures of both components. A comprehensive theoretical
description of this process is constructed. The equations of
evolution of the vapour layer and temperature fields in liquids
within the vicinity of the layer are obtained. The
growing-vapour-layer solution to these equations is derived. The
vapour layer breakaway due to its buoyancy and consequent vapour
bubble formation are described, and the relationships between
macroscopic parameters of the boiling system state and the derived
solution are established for the cases of a well-stirred system
and a stratified system.

The process parameters are evaluated for realistic systems, such
as the n-heptane--water one. The relevance of the case we
considered is revealed for combustion of a light inflammable
liquid over a heavy nonflammable one and demonstrated
experimentally for n-heptane--water and ``white spirit''--water
systems. The theory based results are found to match well the
experimental observations for the n-heptane--water system.

The auxiliary problem of the instability of a thin horizontal
vapour layer between two liquids to bubble formation has been
solved (Appendix~\ref{sec:app:instability}). This solution
provides information required for calculation of the
characteristic size of bubbles, spatial density of bubble
formation centers on the interface, and limitation on the vapour
layer thickness which can be attained before the breakaway of
vapour layer.

Remarkably, for the problem of the bulk boiling the key question
is the rate of nucleation. The answering to this question on the
basis of the theoretical consideration without employment of
semi-empiric information is a challenging task heavily requiring
approaches from the statistical physics theory of nonequilibrium
systems~\cite{Pitaevskii-Lifshitz-v10} and, in particular, the
theory of hydrodynamic fluctuations~\cite{Pitaevskii-Lifshitz-v9}.
On the contrast, the theory of boiling of system of immiscible
liquids below their bulk boiling points can be constructed from
scratch on the mere basis of the macroscopic fluid dynamics.

\begin{acknowledgement}
We are grateful to Prof.\ Alexander N.\ Gorban for provoking the
interest to this problem and Dr.\ Sergey V.\ Shklyaev for fruitful
discussions. We thank our colleagues from the Laboratory of
Hydrodynamic Stability of the Institute of Continuous Media
Mechanics in Perm for help with the demonstration experiments for
the n-heptane--water system; they are Dr.\ Alexey I.\ Mizev,
Prof.\ Konstantin G.\ Kostarev, and Dr.\ Andrey V.\ Shmyrov. The
work has been financially supported by the Russian Science
Foundation grant no.\ 14-21-00090.
\end{acknowledgement}

\appendix

\section{Distribution of species in the vapour layer}
\label{sec:app:distribution}
In this appendix section we derive the particle number density
distribution within a vapour layer linearly growing with time, and
demonstrate it to be of nearly linear profile, see
Eqs.~(\ref{eq-19})--(\ref{eq-20}).

For the idealised symmetric case the problem was found to have an exact solution of the form $L\propto t$, $\Theta\propto t$ with linear profile of the distribution of the particle number densities~\cite{Pimenova-Goldobin-JETP-2014}. We expect a ``successor'' of this solution to exist for an asymmetric case. Accordingly, let us seek the solutions to Eqs.~(\ref{eq-04}) with boundary conditions (\ref{eq-05})--(\ref{eq-08}) in the form of a series in polynomials of $z$;
\begin{align}
n_1(z,t)=n_{1\ast}^{(0)}-\gamma_2\Theta(t)+\alpha\big(z+L_2(t)\big)
\qquad
\nonumber\\[5pt]
 {}+\beta\big(z-L_1(t)\big)\big(z+L_2(t)\big)+\dots\,,
\label{eq-apa-01}
\\[5pt]
n_2(z,t)=n_{2\ast}^{(0)}-\gamma_1\Theta(t)-\alpha\big(z-L_1(t)\big)
\qquad
\nonumber\\[5pt]
 {}-\beta\big(z-L_1(t)\big)\big(z+L_2(t)\big)+\dots\,,
\label{eq-apa-02}
\end{align}
and see whether the terms in these series are proportional to
powers of a small parameter, allowing one to neglect all terms but
the leading ones which are linear functions of $z$. The quadratic
in $z$ terms, with coefficient $\beta$, are intentionally
constructed so that they vanish at the layer boundaries. Here the
boundary conditions (\ref{eq-05})--(\ref{eq-06}) require
\begin{equation}
\alpha=(\gamma_1+\gamma_2)\frac{\Theta}{L}\,,
\label{eq-apa-03}
\end{equation}
and, for $\alpha$ constant in time with linearly growing $\Theta$
and $L$, Eqs.~(\ref{eq-04}) yield
\begin{align}
-\gamma_2\dot{\Theta}+\alpha\dot{L}_2&=2D_{12}\beta\,,
\label{eq-apa-04}
\\
-\gamma_1\dot{\Theta}+\alpha\dot{L}_1&=-2D_{12}\beta\,.
\label{eq-apa-05}
\end{align}

On the other hand, Eqs.~(\ref{eq-09}) and (\ref{eq-10}) yield
\begin{align}
\dot{L}_1&=\frac{D_{12}}{n_{2\ast}^{(0)}}(\alpha+\beta L)\,,
\label{eq-apa-06}
\\
\dot{L}_2&=\frac{D_{12}}{n_{1\ast}^{(0)}}(\alpha-\beta L)\,.
\label{eq-apa-07}
\end{align}

Substituting $\Theta=(\gamma_1+\gamma_2)^{-1}\alpha L$ from
Eq.~(\ref{eq-apa-03}) and $\dot{L}_i$ from (\ref{eq-apa-06}) and
(\ref{eq-apa-07}) into equation system
(\ref{eq-apa-04})--(\ref{eq-apa-05}), one can obtain
\begin{align}
\beta=\frac{\alpha^2}{4}\frac{n_{0\ast}}{n_{1\ast}^{(0)}n_{2\ast}^{(0)}}
 \left(\frac{n_{2\ast}^{(0)}-n_{1\ast}^{(0)}}{n_{2\ast}^{(0)}+n_{1\ast}^{(0)}}
 -\frac{\gamma_2-\gamma_1}{\gamma_2+\gamma_1}\right)\quad
 \nonumber\\[5pt]
 \times\left[1+\mathcal{O}\left(\frac{\beta L}{\alpha}\right)\right].
\label{eq-apa-08}
\end{align}
With Eq.~(\ref{eq-apa-08}), one can see
$$
\frac{\beta L}{\alpha}\sim\frac{\alpha L}{4n_{0\ast}}
 =\frac{(\gamma_1+\gamma_2)\Theta}{4n_{0\ast}}
$$
which is small as required for the series (\ref{eq-apa-01}) and
(\ref{eq-apa-02}) to be series in a small parameter
$(\gamma_1+\gamma_2)\Theta/n_{0\ast}$. The cubic in $z$ term in
series (\ref{eq-apa-01}) and  (\ref{eq-apa-02}) can be further
demonstrated to be small compared to the quadratic term. Thus, for
the leading order of accuracy, it is enough to keep the linear in
$z$ terms in Eqs.~(\ref{eq-apa-01}) and (\ref{eq-apa-02}) and
neglect the quadratic and higher ones.

Notice, according to Eq.~(\ref{eq-apa-08}), $\beta$ exactly
vanishes for $n_{2\ast}^{(0)}/n_{1\ast}^{(0)}=\gamma_2/\gamma_1$
and the linear profile solution becomes an exact one. More
generally, for the special case of
$n_{2}^{(0)}(T)/n_{1}^{(0)}(T)=const$, the linear-profile solution
is exact for arbitrary overheating.

\section{Assessment of the dependence of $(\delta{S}/\delta{V})$ on the heat influx for a well-stirred system}
\label{sec:app:dSdV}
In this appendix section we attempt to derive the rough relationships between the macroscopic parameter $(\delta{S}/\delta{V})$ of the system state and the heat influx rate per unit volume $\dot{Q}_V=\delta{Q}/(\delta{V}\delta{t})$ for a statistically stationary process of interfacial boiling.

The flow and consequent stirring in the system are enforced by the buoyancy of the vapour bubbles, while other mechanisms counteract the stirring of the system. These other mechanisms are gravitational stratification of two liquids, surface tension tending to minimise the interface area and viscous dissipation of the flow energy. Since the latent heat of phase transitions and heat of temperature inhomogeneities are enormously large compared to the realistic values of the kinetic energy of microscopic motion and gravitational potential energy[\footnote{Indeed, the energy of thermal motion of atoms corresponds to characteristic atom velocities $10^2-10^3\,\mathrm{m/s}$, while nothing comparable can be imagined for macroscopic flow velocities in realistic situations. The latent heat of water evaporation is even significantly bigger than the kinetic energy of thermal motion of its atoms at $T=300\,\mathrm{K}$.}], the latter can be neglected in consideration of the heat balance. Hence, all the heat inflow into the system can be considered to be spent for the vapour generation;
 $\dot{Q}_VV\longrightarrow(\Lambda_1n_{1\ast}^{(0)}+\Lambda_2n_{2\ast}^{(0)})\dot{V}_v$,
where $V$ is the system volume, and $\dot{V}_v$ is the volume of the vapour produced in the system per unit time. Thus,
\begin{align}
\dot{V}_v=\frac{\dot{Q}_V\,V}
 {\Lambda_1n_{1\ast}^{(0)}+\Lambda_2n_{2\ast}^{(0)}}\,.
\label{eq-apb-01}
\end{align}

The potential energy of buoyancy of rising vapour bubbles $\rho_lV_vgh/2$ (where $h$ is the linear size of the system, $h\sim V^{1/3}$, $\rho_l$ is the average density of liquids, the vapour density is zero compared to the liquid density) is converted into the kinetic energy of liquid flow, the potential energy of a stirred state of the two-liquid system, the surface tension energy and dissipated by viscosity forces. In a statistically stationary state, the mechanical kinetic and potential energies do not change averagely over time and all the energy influx is to be dissipated by viscosity;
$$
\rho_lV_vgh/2\longrightarrow\dot{W}_{l,k}\tau\,,
$$
where $\dot{W}_{l,k}$ is the rate of viscous dissipation of energy, $\tau$ is the time of generation of the vapour volume $V_v$, $V_v=\dot{V}_v\tau$. Hence,
\begin{align}
\rho_l\dot{V}_vg\frac{h}{2}\sim\dot{W}_{l,k}\,.
\label{eq-apb-02}
\end{align}

Let us estimate the viscous dissipation of the kinetic energy of flow $W_{l,k}$;
\begin{align}
\dot{W}_{l,k}&=\int\limits_V\vec{v}\cdot\vec{f}_\mathrm{vis}\mathrm{d}V
 \sim\int\limits_V\vec{v}\cdot\left(-\eta_l\frac{\vec{v}}{H^2}\right)\mathrm{d}V
\nonumber\\
 \sim& -\frac{\eta_l}{\rho_l}
 \frac{2}{\left(\frac{H_1+H_2}{2}\right)^2}
 \int\limits_V\frac{\rho_lv^2}{2}\mathrm{d}V
 \sim-8\nu_l\left(\frac{\delta{S}}{\delta{V}}\right)^2W_{l,k}\,.
\label{eq-apb-03}
\end{align}
Here $\vec{v}$ is the liquid velocity, $\vec{f}_\mathrm{vis}$ is the viscous force per unit volume, $H$ is the spatial scale of flow inhomogeneity, which is the half-distance between the sheets of the folded interface between liquid components, $\eta_l$ and $\nu_l$ are the characteristic dynamic and kinematic viscosities of liquids, respectively.

Further, we have to establish the relationship between the flow kinetic energy and the mechanical potential energy in the system. Rising vapour bubbles pump the mechanical energy into the system, while its stochastic dynamics is governed by interplay of its flow momentum and the forces of the gravity and the surface tension on the interface. In thermodynamic equilibrium, the total energy is strictly equally distributed between potential and kinetic energies related to quadratic terms in Hamiltonian (this statement is frequently simplified to a less accurate statement, that energy is equally distributed between kinetic and potential energies associated with each degree of freedom). Being not exactly in the case where one can rigorously speak of thermalization of the stochastic Hamiltonian system dynamics, we still may assess the kinetic energy of flow to be of the same order of magnitude as the mechanical potential energy of the system. Thus,
\begin{align}
W_{l,k}\sim W_{l,pg}+W_{l,p\sigma}\,,
\label{eq-apb-04}
\end{align}
where $W_{l,pg}$ and $W_{l,p\sigma}$ are the gravitational potential energy and the surface tension energy, respectively. We set the zero levels of these potential energies at the stratified state of the system with a flat horizontal interface.

The gravitational potential energy of the well-stirred state with uniform distribution of two phases over hight is
$$
W_{l,pg}\sim\Delta\rho_lVg\frac{h}{2}\,,
$$
where $\Delta\rho_l$ is the component density difference.
The surface tension energy is
$$
W_{l,p\sigma}\sim(\sigma_1+\sigma_2)V\left(\frac{\delta{S}}{\delta{V}}\right)\,,
$$
where we neglected the interface area of the stratified state compared to the area $V(\delta{S}/\delta{V})$ in the well-stirred state. Due to the presence of the vapour layer between liquids the effective surface tension coefficient of the interface is $(\sigma_1+\sigma_2)$ but not $\sigma_{12}$ as it would be in the absence of the vapour layer.

Collecting Eqs.~(\ref{eq-apb-01})--(\ref{eq-apb-04}), one finds
\begin{align}
&
\rho_l\frac{\dot{Q}_V\,V}{\Lambda_1n_{1\ast}^{(0)}+\Lambda_2n_{2\ast}^{(0)}}g\frac{h}{2}
\nonumber\\
&\quad
 \approx
 8\nu_l\left(\frac{\delta{S}}{\delta{V}}\right)^2
 \left[\Delta\rho_lVg\frac{h}{2}+(\sigma_1+\sigma_2)V\left(\frac{\delta{S}}{\delta{V}}\right)\right]\,.
\nonumber
\end{align}
This equation can be simplified to
\begin{align}
\dot{Q}_V\approx
 B\left(\frac{\delta{S}}{\delta{V}}\right)^2
 \left[1+\frac{2}{k_{12}^2h}\left(\frac{\delta{S}}{\delta{V}}\right)\right]\,,
\label{eq-apb-05}
\end{align}
where $B=8\nu_l(\Lambda_1n_{1\ast}^{(0)}+\Lambda_2n_{2\ast}^{(0)})\Delta\rho_l/\rho_l$ and $k_{12}$ is given by Eq.~(\ref{eq-apc-21}). Noteworthy, the relative importance of the first and second terms in the brackets in Eq.~(\ref{eq-apb-05}) depends on the system size $h$.

For the n-heptane--water system, $B\approx1.5\,\mathrm{J/(m\cdot s)}$ and $l_{k_{12}}\equiv1/k_{12}\approx0.5\,\mathrm{cm}$. For a well-stirred system the distance between sheets of the folded interface $(\delta{S}/\delta{V})^{-1}\ll h$. The average compound of these two values can be either small or large compared to $l_{k_{12}}$;
\\
(1)~$h\cdot(\delta{S}/\delta{V})^{-1}\ll l_{k_{12}}^2$ corresponds to the case of the surface tension dominated system,
\\
(2)~$h\cdot(\delta{S}/\delta{V})^{-1}\gg l_{k_{12}}^2$ corresponds to the case of a gravity-driven system.

Cubic equation~(\ref{eq-apb-05}) possesses only one positive solution which is real-valued for any value of $\dot{Q}_V/B$;
\begin{align}
\left(\frac{\delta{S}}{\delta{V}}\right)
 =\left(\frac{\delta{S}}{\delta{V}}\right)_g\cdot
 G\left(\frac{\big(\dot{Q}_V/B\big)^{1/2}}{k_{12}^2V^{1/3}}\right),
\label{eq-apb-06}
\end{align}
where $(\delta{S}/\delta{V})_g=(\dot{Q}_V/B)^{1/2}$ is the value of parameter $(\delta{S}/\delta{V})$ for a gravity-driven system and function $G(s)=(6s)^{-1}(R+R^{-1}-1)$, $R=(\sqrt{27}s+\sqrt{27s^2-1})^{2/3}$; $G(0)=1$ and $G(s\gg1)=(2s)^{-1/3}$.

Expression~(\ref{eq-apb-06}) allows estimating the value of parameter $(\delta{S}/\delta{V})$ as a function of heat influx $\dot{Q}_V$ to the system.

\section{Gravitational instability of the vapour layer in stratified system}
\label{sec:app:instability}
In this appendix section we discuss the scenario of vapour layer breakaway for the case of the stratified system as in demonstration experiment in Fig.~\ref{fig2}. In this case the breakaway of the vapour layer is related to a kind of Rayleigh--Taylor instability~\cite{Rayleigh-1883,Taylor-1950} of the upper liquid--vapour interface, where a heavy liquid lies above a nearly weightless fluid. However, our case is untypical, as we are interested specifically in the case of a gas layer between two liquids, and, which is more peculiar, this layer is extremely thin for the situations of our interest (in the next paragraph we will estimate the characteristic thickness of the vapour layer). Practically, this problem would even not arise without the process of vapour layer formation on the two-liquid contact interface, as there seems to be no other robust mechanism of appearance and persistent maintenance of such a thin layer. To the authors' knowledge, this problem is not addressed in the literature.

In order to estimate the reference thickness of the layer, one can
look at Fig.~\ref{fig2}. In Fig.~\ref{fig2}(b),
the typical diameter of vapour bubbles detaching the interface is
$d_b\sim1\,\mathrm{mm}$ and the bubble lanes stand at
characteristic distance of $1\,\mathrm{cm}$ from each other, i.e.,
each bubble is formed by the vapour layer patch breaking-away from
the interface area $S_{vl}\sim1\,\mathrm{cm}^2$. The bubble volume
$V_b=(\pi/6)d_b^3$ is equal to the layer patch volume
$h_{vl}S_{vl}$; therefore, the characteristic thickness of the
vapour layer $h_{vl}=V_b/S_{vl}\sim10^{-5}\mathrm{m}$.

All the consideration in this section in focused on the specific
fluid dynamical problem and only the final results are employed in
the main paper. Thus, for the convenience reason, in this section
we will use notations independent of the notations in the main
paper.

\begin{figure}[t]
\center{
\includegraphics[width=0.3\textwidth]%
{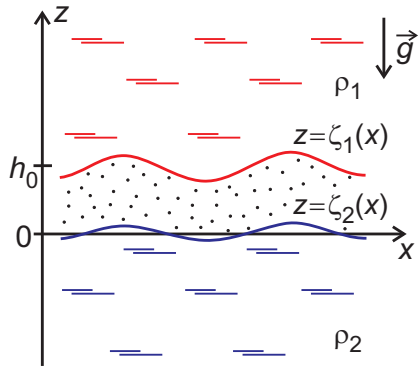}}
\caption{
System of two liquids with densities $\rho_1$ and $\rho_2>\rho_1$
separated by a thin layer of viscous gas.
 }
\label{fig5}
\end{figure}

We consider the gravity-capillary waves and the system instability
to their growth. For the linear stability analysis it is enough to
consider plane-wave perturbations, i.e., the problem can be
investigated in the $(x,z)$-geometry, where $z$ is the vertical
coordinate and $x$ is the coordinate along the wave vector (see
Fig.~\ref{fig5}). We consider thicknesses of liquid layers to be
large compared to the interface inflection wavelength, in which
case one can assume the unperturbed liquids to occupy half-spaces
$z>h_0$ and $z<0$. The densities of the upper light liquid and the
lower heavy liquid are $\rho_1$ and $\rho_2>\rho_1$, respectively,
and the vapour is nearly weightless. The positions of the
vapour-liquid interfaces are $z=\zeta_1(x,t)$ and $z=\zeta_2(x,t)$
(Fig.~\ref{fig5}); for the unperturbed state $\zeta_1=h_0$ and
$\zeta_2=0$, where $h_0$ is the unperturbed vapour layer
thickness.

The classical marginal Rayleigh--Taylor instability (with no
vapour layer) is monotonous and, therefore, the instability
threshold is unaffected by viscosity. The problem can be
considered for inviscid liquids. The density of the vapour between
liquids is nearly zero compared to the liquid densities and its
flow should be inertialess. On the other hand, the difference
$\dot{\zeta}_1-\dot{\zeta}_2=\dot{h}$ forces vapour redistribution
and can create a strong flow in a narrow gap, which, in the
absence of inertia, is restrained only by viscosity. Hence, the
viscosity has to be accounted for the vapour layer. Characteristic
hydrostatic and hydrodynamic pressure gradients in the system are
small compared to atmospheric pressure and one can treat the
vapour to be incompressible. Summarizing, we deal with a system
where the liquid flow can be assumed inviscid while the vapour
flow is incompressible and dominantly viscous, its velocity is
much larger than that of liquids.

Let us consider flow in the vapour layer. The layer thickness
$h=\zeta_1-\zeta_2$ is small compared to the characteristic
horizontal scale of the wave pattern and hence the flow is nearly
parallel to the layer middle surface and pressure is nearly
constant across the layer, $p_g(x,z)=p_g(x)$. Additionally, since
the vapour flow is much faster that the liquid flows, one can
adopt zero-velocity boundary conditions for the vapour. Thus, one
can find the vapour viscous flow to be a Poiseuille flow along a
thin gap between two planes:
\begin{equation}
v(x,z)=v_m(x)\left(1-\frac{4(z-\zeta_m)^2}{h^2}\right)\,,
\label{eq-apc-01}
\end{equation}
where $v$ is the vapour velocity tangential to the layer middle
surface, $\zeta_m=(\zeta_1+\zeta_2)/2$ is the $z$-coordinate of
the layer middle surface, $v_m$ is the vapour velocity at
$\zeta_m$. The Navier-Stokes equation for vanishing fluid density
provides relation between the flow and the pressure gradient;
\begin{equation}
\frac{\partial}{\partial x}p_g(x,t)=\eta\frac{\partial^2v}{\partial z^2}
 =-\frac{8\eta}{h^2}v_m(x,t)\,,
\label{eq-apc-02}
\end{equation}
where $p_g$ is pressure within the layer, $\eta$ is the dynamic
viscosity. The layer thickness change $\dot{h}$ is owned by the
fluid flux through the layer cross-section
$Q=\int_{\zeta_1}^{\zeta_2}v\,\mathrm{d}z=(2/3)v_mh$;
$\dot{h}=-(\partial/\partial x)Q$. Using the relation between
$v_m$ and pressure gradient~(\ref{eq-apc-02}), one can write
$$
\dot{h}=\frac{\partial}{\partial x}
 \left(\frac{h^3}{12\eta}\frac{\partial p_g}{\partial x}\right).
$$
Since we are to consider infinitesimal perturbations of the state
with flat interfaces, we need only contributions which are linear
in small parameters $(p_g-p_{g0})$ and $(h-h_0)$;
\begin{equation}
\dot{h}=\frac{h_0^3}{12\eta}\frac{\partial^2p_g}{\partial x^2}.
\label{eq-apc-03}
\end{equation}

Let us now consider the mechanics of liquid phases. Considering
inviscid flow, one can describe the current velocity with the
stream scalar potential $\Phi_j$; the $j$-th liquid velocity
$\vec{v}_j=\nabla\Phi_j$. The incompressibility condition
$$
\nabla\cdot\vec{v}_j=0
$$
requires $\Phi_j$ to be harmonic functions;
\begin{equation}
\mathrm{\Delta}\Phi_j(x,z,t)=0\,.
\label{eq-apc-04}
\end{equation}
The Euler equation in terms of potential takes the form
$$
\rho_j\nabla\left(\dot{\Phi}_j+\frac{1}{2}(\nabla\Phi_j)^2\right)
 =-\nabla(p_j+\rho_jgz)\,,
$$
where $p_j$ is the pressure field in the $j$-th liquid, $g$ is the
gravity. Thus, one can evaluate the pressure field for a given
flow;
\begin{equation}
p_j=p_{j,0}-\rho_j\dot{\Phi}_j-\rho_jgz\,,
\label{eq-apc-05}
\end{equation}
where the quadratic term is neglected because we consider an
infinitesimal perturbation flow.

The condition of stress balance at the liquid surfaces relates the
pressure jump across the surface with the capillary pressure;
\begin{align}
(p_1-p_g)|_{z=\zeta_1}&
 =\sigma_1\frac{\partial^2\zeta_1}{\partial x^2}\,,
\label{eq-apc-06}\\
(p_2-p_g)|_{z=\zeta_2}&
 =-\sigma_2\frac{\partial^2\zeta_2}{\partial x^2}\,,
\label{eq-apc-07}
\end{align}
where $\sigma_j$ is the surface tension of the $j$-th liquid.

Since we treat the stability of the flat-interface state and
consider small perturbations, it is convenient to formulate
equations in domains $z<0$ and $z>h_0$ rather than in domains
$z<\zeta_2$ and $z>\zeta_1$. Consequently, the boundary conditions
should be moved to $z=0$ and $z=h_0$ from $z=\zeta_2$ and
$z=\zeta_1$, respectively. To do so for boundary conditions
(\ref{eq-apc-06})--(\ref{eq-apc-07}), one has to employ
Eq.~(\ref{eq-apc-05}). Up to the linear in perturbation terms,
boundary conditions (\ref{eq-apc-06})--(\ref{eq-apc-07}) can be
recast into the following form:
\begin{align}
-\rho_1\dot\Phi_1|_{z=h_0}-p_g-\rho_1g(\zeta_1-h_0)
 =\sigma_1\frac{\partial^2\zeta_1}{\partial x^2}\,,
\label{eq-apc-08}\\
-\rho_2\dot\Phi_2|_{z=0}-p_g-\rho_2g\zeta_2
 =-\sigma_2\frac{\partial^2\zeta_2}{\partial x^2}\,.
\label{eq-apc-09}
\end{align}

The kinematic boundary conditions remain to be accounted for. The
liquid flow shifts the liquid--vapour interface; $\dot{\zeta}_j$
is determined by the $z$-component of $j$-th liquid velocity on
the liquid surface, in terms of potential,
\begin{align}
\dot{\zeta_1}&=\left.\frac{\partial \Phi_1}{\partial z}\right|_{z=h_0},
\label{eq-apc-10}
\\
\dot{\zeta_2}&=\left.\frac{\partial \Phi_2}{\partial z}\right|_{z=0}.
\label{eq-apc-11}
\end{align}
Substituting the thickness variation
$\dot{h}=\dot{\zeta_1}-\dot{\zeta_2}$ in Eq.~(\ref{eq-apc-03}), one
finds
\begin{equation}
\dot{\zeta_1}-\dot{\zeta_2}
 =\frac{h_0^3}{12\eta}\frac{\partial^2p_g}{\partial x^2}.
\label{eq-apc-12}
\end{equation}

Eqs.~(\ref{eq-apc-04}), (\ref{eq-apc-08})--(\ref{eq-apc-12}) form a
complete system of differential equations for fields
$\Phi_j(x,z,t)$, $\zeta_j(x,t)$ and $p_g(x,t)$. As the equations
for perturbations are homogeneous in the $x$-direction and in
time, the solution can be sought in a normal form $\propto
e^{\lambda t+ikx}$. For $\Phi_j\propto e^{ikx}$,
Eq.~(\ref{eq-apc-04}) yields $\Phi_1\propto e^{-kz+ikx}$ and
$\Phi_2\propto e^{kz+ikx}$. Hence, Eqs.~(\ref{eq-apc-10}) and
(\ref{eq-apc-11}) yield $\Phi_1|_{z=h_0}=-k^{-1}\dot{\zeta}_1$ and
$\Phi_2|_{z=0}=k^{-1}\dot{\zeta}_2$, respectively. With
substitution of $\Phi_j$ and normal perturbation
$$
\{\zeta_1,\zeta_2-h_0,p_g\}=\{\xi_1,\xi_2,P\}\,e^{\lambda t+ikx},
$$
Eqs.~(\ref{eq-apc-08}), (\ref{eq-apc-09}) and (\ref{eq-apc-12}) can
be cast into the following equation system for amplitudes $\xi_1$,
$\xi_2$, $P$:
\begin{align}
 \left(\rho_1\frac{\lambda^2}{k}
  -\rho_1g+\sigma_1k^2\right)\xi_1-P&=0\,,
\label{eq-apc-13}\\
 \left(-\rho_2\frac{\lambda^2}{k}
  -\rho_2g-\sigma_2k^2\right)\xi_2-P&=0\,,
\label{eq-apc-14}\\
 \lambda\xi_1-\lambda\xi_2+\frac{h_0^3}{12\eta}k^2P&=0\,.
\label{eq-apc-15}
\end{align}
The corresponding characteristic equation for the exponential
growth rate $\lambda$ reads
\begin{align}
\nonumber
&\lambda\left(\lambda^2+\frac{\rho_2-\rho_1}{\rho_2+\rho_1}gk
 +\frac{\sigma_1+\sigma_2}{\rho_2+\rho_1}k^3\right)\\[5pt]
&\quad
 {}+Ak\left(\lambda^2-gk+\frac{\sigma_1}{\rho_1}k^3\right)
 \left(\lambda^2+gk+\frac{\sigma_2}{\rho_2}k^3\right)=0\,,
\label{eq-apc-16}
\end{align}
where
$$
A=\frac{h_0^3}{12\eta}\frac{\rho_1\rho_2}{\rho_1+\rho_2}
$$
is the parameter characterising the thickness of the vapour layer.

\paragraph{Result validation: Limiting cases.}
One can consider two limiting cases for the system: thin layer
($h_0\to 0$) and thick layer ($h_0\to\infty$). In the first case
the second term in Eq.~(\ref{eq-apc-16}) can be neglected, and the
equation yields
\begin{align}
\lambda&=0\,,
\label{eq-apc-17}\\
\lambda^2&=-\frac{\rho_2-\rho_1}{\rho_2+\rho_1}gk
 -\frac{\sigma_1+\sigma_2}{\rho_2+\rho_1}k^3\,.
\label{eq-apc-18}
\end{align}
Eq.~(\ref{eq-apc-18}) is exactly the result for gravity-capillary
waves for two-liquid system without vapour layer (however, with
surface tension $\sigma=\sigma_1+\sigma_2$) well known in the
literature (e.g.,~\cite{Sharp-1984}). Since the right part of this
equation is strictly nonpositive, all exponential growth rates
$\lambda$ are imaginary, there is no instability.
Eq.~(\ref{eq-apc-17}) represents the fact that for vanishing $h_0$
the inhomogeneities of pressure in the vapour layer dissolves
infinitely slowly (cf.\ Eq.~(\ref{eq-apc-03})) due to diminished
viscous flow along vanishingly narrow gap. These pressure $p_g$
perturbations form a neutral mode with $\lambda\to0$ for
$h_0\to0$.

In the opposite limiting case, when thickness $h_0$ and parameter
$A$ are large enough, the first term in Eq.~(\ref{eq-apc-16}) can
be neglected and the characteristic equation can be factorised
yielding two independent pairs of solutions:
\begin{align}
\lambda_{1,2}^2&=kg-\frac{\sigma_1}{\rho_1}k^3\,,
\label{eq-apc-19}\\
\lambda_{3,4}^2&=-kg-\frac{\sigma_2}{\rho_2}k^3\,.
\label{eq-apc-20}
\end{align}
These solutions correspond to the case of conventional
gravity-capillary waves on the liquid--gas interface (cf.\
Eq.~(\ref{eq-apc-18})) for the unstable state of the liquid layer
over gas, Eq.~(\ref{eq-apc-19}), and for the stable state of gas
over liquid, Eq.~(\ref{eq-apc-20}). In this case, vapour layer is
thick enough to make the liquid surfaces insensitive to motion of
each other. Eq.~(\ref{eq-apc-20}) has only imaginary solutions, as
it should be, while Eq.~(\ref{eq-apc-19}) has a pair of real roots
for $k<k_1$, where $k_1=\sqrt{\rho_1g/\sigma_1}$. Perturbations
with positive $\lambda$ grow exponentially, meaning the system
is unstable.


It is convenient to introduce reference values of the wavenumber:
\begin{equation}
k_1=\sqrt{\frac{\rho_1g}{\sigma_1}}\,,\quad
k_2=\sqrt{\frac{\rho_2g}{\sigma_2}}\,,\quad
k_{12}=\sqrt{\frac{(\rho_2-\rho_1)g}{\sigma_1+\sigma_2}}\,.
\label{eq-apc-21}
\end{equation}

\paragraph{The case of small non-zero $h_0$.}
Solution $\lambda=0$ for $h_0=0$ given by Eq.~(\ref{eq-apc-17})
requires an additional subtle treatment as it can be made non-zero
and change its sign with arbitrary small corrections. Thus, the case
of small but non-zero $h_0$ may be not represented by
solution~(\ref{eq-apc-17}) for $h_0=0$ well.

Let us first calculate the reference values for $\lambda$, $A$,
and $k$: $\lambda_\ast$, $A_\ast$, and $k_\ast$. One can notice,
that after substitution of $\lambda=0$, Eq.~(\ref{eq-apc-16}) can
be satisfied with $k=k_1$ regardless of the value of $A$, and this
is the only non-zero value of $k$ which satisfies the equation for
$\lambda=0$. Thus $\lambda(k)$ crosses the zero point only at
$k=k_1$. Hence, it is natural to chose
$$
k_\ast=k_1\,.
$$
Further, for $k=k_\ast$ the reference value of the terms near
$\lambda^2$ in brackets of Eq.~(\ref{eq-apc-16}) is $gk_\ast$,
suggesting
$$
\lambda_\ast=\sqrt{gk_1}\,.
$$
The first and second terms in Eq.~(\ref{eq-apc-16}) for $k=k_\ast$
and $\lambda=\lambda_\ast$ are commensurable when $A$ equals the
reference value
$$
A_\ast=\frac{1}{k_\ast\lambda_\ast}=g^{-1/2}k_1^{-3/2}
 =\left(\frac{\sigma_1}{\rho_1}\right)^{3/4}g^{-5/4}\,.
$$
The corresponding reference value of $h_0$ is
\begin{equation}
h_{0\ast}=\left(\frac{12\eta}{g^{5/4}}
 \frac{\rho_1+\rho_2}{\rho_1\rho_2}\right)^{1/3}
 \left(\frac{\sigma_1}{\rho_1}\right)^{1/4}.
\label{eq-apc-22}
\end{equation}

For the n-heptane--water system as an example, with data from
Tab.~\ref{tab1}, one finds $k_1=659\,\mathrm{m^{-1}}$, i.e., the reference instability
wavelength $2\pi/k_1=1.0\,\mathrm{cm}$, $A_\ast=1.9\cdot10^{-5}$, $h_{0\ast}=0.15\,\mathrm{mm}$; $k_2=389\,\mathrm{m^{-1}}$ and $k_{12}=206\,\mathrm{m^{-1}}$. The previously estimated thickness $h_{vl}\sim0.01\,\mathrm{mm}$ of the layer suffering bubble breakaway we observed in the experimental demonstration (Fig.~\ref{fig2}) is by factor $15$ smaller than $h_{0\ast}$, meaning we can reliably restrict our consideration to the case $h_0\ll h_{0\ast}$.

\begin{figure}[t]
\center{
\includegraphics[width=0.35\textwidth]%
{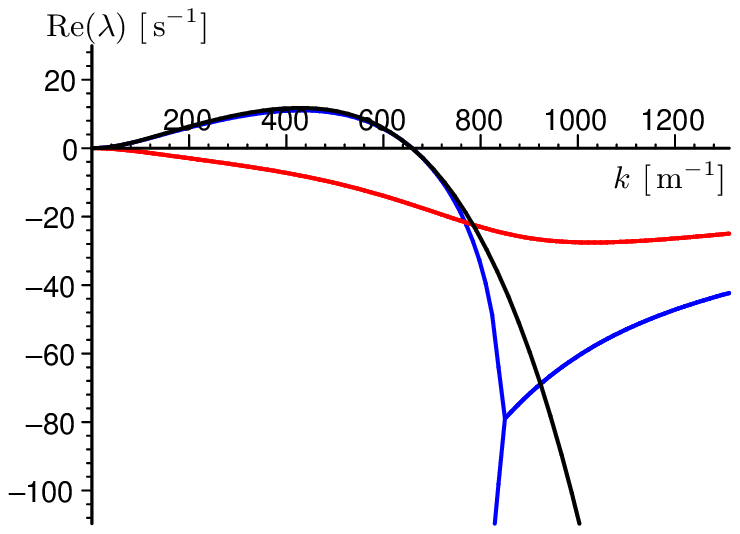}\\[10pt]
\includegraphics[width=0.35\textwidth]%
{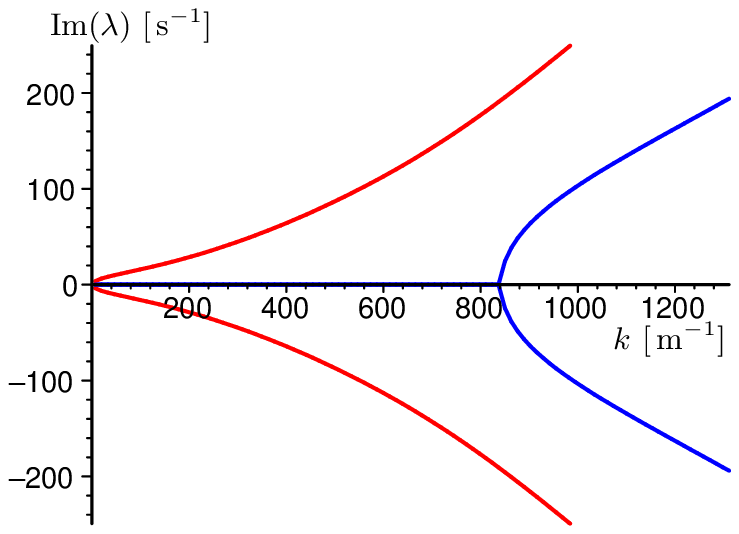}
}
\caption{
Spectrum of the exponential growth rates $\lambda(k)$ for n-heptane--water system with $A=0.3A_\ast$ has 4 branches given by Eq.~(\ref{eq-apc-16}). The pair of $\lambda$ plotted with red lines is a pair of complex conjugated values, $Re(\lambda)$ is the same for both branches. The pair of $\lambda$ plotted with blue lines is complex conjugated where the imaginary part of $\lambda$ is non-zero; otherwise, there is two real-valued branches, one of which can take positive values, always crossing the abscissa at $k=k_1$. The black solid line represents approximation~(\ref{eq-apc-23}) for the only branch of $\lambda$ with a non-negative real part.
 }
\label{fig6}
\end{figure}

For the solution branch $\lambda=0$ distorted by non-zero $A\ll A_\ast$, say $\lambda_1(k)$, $|\lambda_1|\ll\lambda_\ast$ by continuity. Therefore, Eq.~(\ref{eq-apc-16}) turns into
$$
\lambda_1\frac{\sigma_1+\sigma_2}{\rho_2+\rho_1}k(k^2+k_{12}^2)
 +Ak^3\frac{\sigma_1\sigma_2}{\rho_1\rho_2}(k^2-k_1^2)(k^2+k_2^2)=0\,,
$$
whence
\begin{equation}
\lambda_1=\frac{h_0^3}{12\eta}\frac{\sigma_1\sigma_2}{\sigma_1+\sigma_2}
 \frac{k^2(k_1^2-k^2)(k_2^2+k^2)}{k^2+k_{12}^2}\,.
\label{eq-apc-23}
\end{equation}

In Fig.~\ref{fig6} one can see this solution to match the exact solution well even for a non-small $A/A_\ast$.
Although the maximum point of dependence $\lambda_1(k)$ is unique and corresponds to the unique positive solution of the equation
\begin{align}
\frac{\mathrm{d}\lambda_1}{\mathrm{d}(k^2)}=0
 =&2k^6+(3k_{12}^2-k_1^2+k_2^2)k^4
\nonumber\\
 &-2(k_1^2-k_2^2)k_{12}^2 k^2-k_1^2k_2^2k_{12}^2\,,
\nonumber
\end{align}
this analytical solution is too lengthy and simultaneously can be trivially derived. Hence, we omit the general expression and provide the value of $k_\mathrm{max}$ specific to the n-heptane--water system;
$k_{\mathrm{max}}^{n\mathrm{C_7H_{16}-H_2O}}=429\,\mathrm{m^{-1}}$. Thus,
\begin{equation}
\lambda_\mathrm{max}^{n\mathrm{C_7H_{16}-H_2O}}
=1.07\cdot 10^{13}\mathrm{m^{-3}s^{-1}}\times h_0^3\,.
\label{eq-apc-24}
\end{equation}
For approximate calculations one can avoid solving the equation for $k_\mathrm{max}$ and use the value $k_{1\mathrm{max}}=k_1/\sqrt{3}$, which maximizes expression~(\ref{eq-apc-19}); $k_{1\mathrm{max}}$ is always close to $k_\mathrm{max}$. The analytical assessment of the exponential growth rate reads then
\begin{equation}
\lambda_1(k_{1\mathrm{max}})=\frac{h_0^3}{54\eta}\frac{\sigma_1\sigma_2}{\sigma_1+\sigma_2}
 \frac{k_1^4(k_2^2+k_1^2/3)}{k_{12}^2+k_1^2/3}\,.
\label{eq-apc-25}
\end{equation}
For the n-heptane--water system the last expression yields $1.10\cdot 10^{13}\mathrm{m^{-3}}\times h_0^3$ which is only $3\%$ larger than expression~(\ref{eq-apc-24}) and thus can be treated as a generally reliable assessment.

\section{Calculation of physical parameters of the vapour mixture}
\label{sec:app:params}
\subsection{Saturated vapour number density and the interfacial boiling point $T_\ast$}
The experimental data on the dependence of saturated vapour pressure (or particle number density) on temperature for some substances may be lacking, not enough detailed, or not easily accessible in the literature. Under such circumstances one can use a straightforward theoretical approximation (Eq.~(B13) in Appendix B of~\cite{Goldobin-Brilliantov-2011}), which proved to work well for water vapour; for n-heptane it yields results well matching the independent experimental data on the enthalpy of evaporation and saturated vapour pressure at the standard conditions. Assuming the vapour to be a perfect gas and the liquid phase to have temperature-independent thermodynamics properties (they are temperature-independent within the temperature range of our interest), one can find (e.g., see Appendix B of~\cite{Goldobin-Brilliantov-2011}) the ratio of the saturated vapour pressure to pressure (or $n^{(0)}(T,P)/n_0$)
\begin{eqnarray}
Y=\frac{n^{(0)}}{n_0}=
 \frac{P_0}{P}
 \left(\frac{T}{T_0}\right)^{\textstyle\frac{\Delta c_p}{k_\mathrm{B}}}
 \exp\left[\frac{v_\mathrm{liq}(P-P_0)}{RT}
 \right.
\nonumber\\[5pt]
 \left.
 {}-\frac{\Delta H_0-\Delta c_p\,T_0}{k_\mathrm{B}}
 \left(\frac{1}{T}-\frac{1}{T_0}\right)\right],\quad
\label{eq-apd-01}
\end{eqnarray}
where $k_B$ is the Boltzmann constant, $R$ is the universal gas constant,
 $\Delta c_p=c_{p,\mathrm{vap}}-c_{p,\mathrm{liq}}$
is the difference between specific heats per one molecule in the vapour and liquid phases, $\Delta H_0$ is the enthalpy of vaporization per one molecule, subscript ``$0$'' indicates the values corresponding to the bulk boiling point $T_0$ at pressure $P_0$, $v_{\mathrm{liq}}$ is the molar volume of the liquid phase. Specifically in the case of our interest, $P=P_0$ and
\begin{equation}
Y(T)=\left(\frac{T}{T_0}\right)^{\textstyle\frac{\Delta c_p}{k_\mathrm{B}}}
 \exp\left[-\frac{\Delta H_0-\Delta c_p\,T_0}{k_\mathrm{B}}
 \left(\frac{1}{T}-\frac{1}{T_0}\right)\right].\quad
\label{eq-apd-02}
\end{equation}
The derivative of $Y(T)$ with respect to temperature yields the coefficient $\gamma$ for Eq.~(\ref{eq-02});
\begin{equation}
\frac{\gamma}{n_0}
 =\left.\frac{\mathrm{d}Y}{\mathrm{d}T}\right|_{T=T_\ast}
 =\frac{\Delta H_0+\Delta c_p(T_\ast-T_0)}{k_\mathrm{B}T_\ast^2}
 Y(T_\ast)\,.
\label{eq-apd-03}
\end{equation}

\begin{table}[t]
\caption{Molecular properties of water and n-heptane.}
\begin{center}
\begin{tabular}{|c|c|c|}
\hline
 & $\mathrm{H_2O}$ & \mbox{n-heptane}
 \\
\hline
\vspace{-7pt}
 &\qquad\qquad\qquad&\qquad\qquad\qquad
 \\
$\Delta H_0/k_\mathrm{B}$ (K)
 & $4892$ & $3821$ \\[5pt]
$\Delta c_p/k_\mathrm{B}$
 & $-5.00$ & $-7.86$ \\[5pt]
$d$ (\AA)
 & $2.70$ & $6.66$ \\[5pt]
$m$ ($10^{-25}$kg)
 & $0.2992$ & $1.664$ \\[3pt]
\hline
\end{tabular}
\end{center}
\label{tab2}
\end{table}

\begin{figure}[t]
\center{
\includegraphics[width=0.4\textwidth]%
{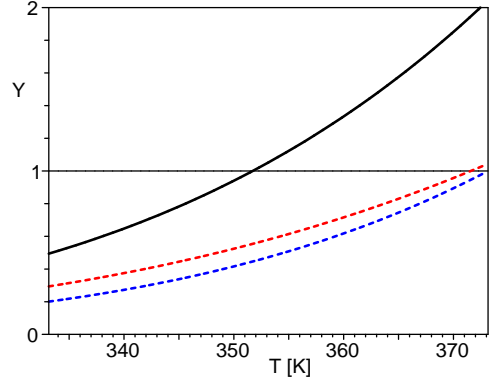}}
\caption{
Relative number densities $Y=n^{(0)}/n_0$ of saturated vapour of water (blue dashed line) and n-heptane (red dashed line) and their sum are plotted vs.\ temperature; $Y(T)$ are calculated with Eq.~(\ref{eq-apd-02}).
 }
\label{fig7}
\end{figure}

With data provided in Tab.~\ref{tab2} and the bulk boiling temperature $T_0$ from Tab.~\ref{tab1}, one can evaluate the saturated vapour density for water and n-heptane. In Fig.~\ref{fig7}, $Y_\mathrm{H_2O}(T)$, $Y_\mathrm{C_7H_{16}}(T)$, and the sum $Y_\mathrm{H_2O}+Y_\mathrm{C_7H_{16}}$ are plotted. The sum attains the value of $1$ at $T_\ast$; numerically solving equation $Y_\mathrm{H_2O}(T_\ast)+Y_\mathrm{C_7H_{16}}(T_\ast)=1$, one finds $T_\ast=351.71\,\mathrm{K}=78.56^\circ\mathrm{C}$, $n_{\mathrm{H_2O}}^{(0)}(T_\ast)=0.446\,n_0$, $n_{\mathrm{C_7H_{16}}}^{(0)}(T_\ast)=0.554\,n_0$, $\gamma_\mathrm{H_2O}/n_0=0.0180\,\mathrm{K^{-1}}$, $\gamma_\mathrm{C_7H_{16}}/n_0=0.0177\,\mathrm{K^{-1}}$.

\subsection{Transport coefficients $D_{12}(T_\ast)$ and $\eta_{12}(T_\ast)$}
For evaluation of the transport coefficients of the vapour mixture the Chapman--Enskog kinetic theory of non-uniform gases~\cite{Chapman-Cowling-1970} can be employed. The first Chapman--Enskog approximation for the diffusion coefficient $D_{12}$ is independent of the component concentration;
\begin{equation}
D_{12}=\frac{3}{8n_0d_{12}^2}\sqrt{\frac{k_\mathrm{B}T}{2\pi}
 \left(\frac{1}{m_1}+\frac{1}{m_2}\right)}\,,
 \quad
 n_0=\frac{P_0}{k_\mathrm{B}T}\,.
\label{eq-apd-04}
\end{equation}
Here $m_j$ is the molecule mass; $\pi d_{12}^2$ is the scattering cross section (for an elastic sphere gas, $d_j$ is the sphere diameter), with a good accuracy $d_{12}=(d_1+d_2)/2$, where $\pi d_j^2$ is the scattering cross section for mutual collisions of the molecules of sort $j$.

According to Wilke~\cite{Wilke-1950}, the ideal gas mixture viscosity can be quite accurately calculated as
\begin{align}
\eta_{12}\approx &
\frac{\eta_1}{\displaystyle
 1+\frac{(n_2/n_1)[1+(\eta_1/\eta_2)^{1/2}(m_2/m_1)^{1/4}]^2}
 {(4/\sqrt{2})[1+(m_1/m_2)]^{1/2}}}
 \nonumber\\[5pt]
 &+\frac{\eta_2}{\displaystyle
 1+\frac{(n_1/n_2)[1+(\eta_2/\eta_1)^{1/2}(m_1/m_2)^{1/4}]^2}
 {(4/\sqrt{2})[1+(m_2/m_1)]^{1/2}}}\,,
\label{eq-apd-05}
\end{align}
where $\eta_j$ is the dynamic viscosity of the pure gas of specie $j$ at atmospheric pressure. At temperature $T_\ast$, $n_j=n_j^{(0)}$. The dynamic viscosity of single component gas can be calculated with the Chapman--Enskog theory; to the forth order~\cite{Chapman-Cowling-1970},
\begin{equation}
\eta_j=1.02513\frac{5}{16}\frac{\sqrt{m_jk_\mathrm{B}T}}
 {\pi^{1/2}d_j^2\Omega_\eta(k_\mathrm{B}T/\varepsilon_{12})}\,,
\label{eq-apd-06}
\end{equation}
where geometric factor $\Omega_\eta$ is of order of $1$ and characterises interparticle interactions during collisions (e.g., for an elastic sphere gas, $\Omega_\eta=1$), $\varepsilon_{12}$ is a reference value of the intermolecular interaction energy.

With Eqs.~(\ref{eq-apd-04})--(\ref{eq-apd-06}) and molecular parameters from Tab.~\ref{tab2}, one can calculate for the n-heptane--water vapour mixture at $T=T_\ast$ ($n_j=n_{j\ast}^{(0)}$ were calculated in the previous subsection): $D_{12}(T_\ast)=1.20\cdot 10^{-5}\mathrm{m^2/s}$ and $\eta_{12}(T_\ast)=0.59\cdot 10^{-5}\mathrm{Pa\cdot s}$.

\end{document}